\documentclass[paper,notoc,nohyper]{JHEP3}
\usepackage{epsfig}
\usepackage{amsbsy} 

\def\MSbar{$\overline{{\rm MS}}$}
\def\e{\epsilon}
\newcommand{\la}{\langle}
\newcommand{\ra}{\rangle}
\def\ket#1{|{#1}\rangle}

\def\P{{\cal P}}

\def\sAB{s_{AB}}
\def\sabc{s_{abc}}
\def\sab{s_{ab}}
\def\sac{s_{ac}}
\def\sbc{s_{bc}}
\def\S{\, {\rm S}}
\def\Split{\, {\rm Split}}
\def\bom#1{{\mbox{\boldmath $#1$}}}
\def\CA{C_A}
\def\CF{C_F}

\def\NF{N_F}
\def\lx{\ln(w)}
\def\ly{\ln(1\!-\!w)}
\def\Libx{{\rm Li}_2(w)}
\def\Licx{{\rm Li}_3(w)}
\def\Lidx{{\rm Li}_4(w)}
\def\Liby{{\rm Li}_2(1\!-\!w)}
\def\Licy{{\rm Li}_3(1\!-\!w)}
\def\Lidy{{\rm Li}_4(1\!-\!w)}
\def\Lidz{{\rm Li}_4\left(\frac{w}{w\!-\!1}\right)}
\def\AAAXXX{\Licx-\zeta_3-\frac{1}{2}\lx\left(\Libx-\zeta_2\right)}
\def\AAAYYY{-\Liby+2\zeta_2}

\def\WW{\frac{-w}{1-w}}
\def\OMWW{\frac{1-w}{-w}}
\def\Li{{\rm Li}}
\def\e{\epsilon}

\title{\boldmath 
Two-loop splitting functions in QCD\footnote{Work supported in part by the UK Particle Physics and Astronomy 
Research Council and by the EU Fifth Framework Programme `Improving Human
Potential', Research Training Network `Particle Physics Phenomenology 
at High Energy Colliders', contract HPRN-CT-2000-00149.}
}
\author{
S.D. Badger  and E.~W.~N.~Glover  \\
Department of Physics,
University of Durham,
Durham DH1 3LE,
England\\
E-mail:  \email{S.D.Badger@durham.ac.uk, E.W.N.Glover@durham.ac.uk}}
\abstract{
We present the universal 
two-loop splitting functions that describe the limits of
two-loop $n$-point amplitudes of massless particles 
when two of the momenta are collinear. 
To derive the splitting amplitudes, we take the collinear limits of explicit
two-loop four-point helicity amplitudes computed 
in the 't Hooft-Veltman scheme.
The $g \to gg$ splitting amplitude has recently been computed using 
the unitarity sewing method and we find complete agreement with the results of
Ref.~\cite{Bern:2lsplit}.   The two-loop $q \to qg $ and $g \to q\bar q$ splitting functions 
are new results.  We also provide an expression for the two-loop soft splitting
function.
}

\preprint{{DCPT/04/42}, {IPPP/04/21}, {hep-ph/0405236}}

\begin{document}

\section{Introduction}
\label{sec:intro}

The singular behaviour of QCD amplitudes is an important ingredient in
understanding the perturbative structure of quantum field theories.   In
general, when one or more final state particles are either soft or collinear,
the amplitudes factorise into the product of an amplitude depending on the
remaining hard partons in the process (including any hard partons constructed
from an ensemble of unresolved partons) and a factor that contains all of the
singularities due to the unresolved particles.   One of the best known
examples of this type of factorisation is the limit of tree amplitudes when
two particles are collinear.  This factorisation is universal and can be
generalised to any number of loops~\cite{Kosower:allorderfact}.

Such factorisation properties play a dual role in developing higher order
perturbative predictions for observable quantities. On the one hand, a
detailed knowledge of the structure of unresolved emission enables phase space
integrations to be organised such that the infrared singularities due to soft
or collinear emission can be analytically 
extracted~\cite{Giele:1992vf,Frixione:1996ms,Catani:1997vz}.   On the other, the
collinear limit plays a crucial role in the unitarity-based method for loop
calculations~\cite{Bern:split1e0,Bern:1995cg,Bern:1996db,Bern:1996je}.

For next-to-leading order (NLO) predictions one of the key ingedients is knowledge of
the the single unresolved configurations where a gluon is soft or where two particles
are collinear. The universal behaviour of tree amplitudes in the single soft
and collinear limits is well known and has been extensively discussed in the
literature (see for example Refs.~\cite{Altarelli:1977zs,Bassetto:1983ik,
Dokshitzer:1991wu,Mangano:multipart,Dixon:scatamp}). Antenna factors that
describe both limits simultaneously have also been
derived~\cite{Kosower:antenna} and subsequently  employed in the construction
of parton level NLO predictions for $e^+e^- \to 4$~jets~\cite{Campbell:4jet}. 
At next-to-next-to-leading order (NNLO) one encounters, for example, tree
amplitudes with double unresolved configurations where two gluons are
soft~\cite{Berends:multigluon,Catani:unpubl} or when three particles are
collinear~\cite{Campbell:dblunres,Catani:NNLOcollfact,Catani:IRtreeNNLO} as
well as the single unresolved limits of one-loop
amplitudes~\cite{Bern:1995ix,Bern:split1e0,Bern:split1gluon,Kosower:split1,Bern:split1QCD,Catani:2000pi}.  
Currently, there is significant effort to extract the infrared singularities
from such
processes~\cite{Kosower:2002su,Kosower:2003cz,Weinzierl:2003fx,Weinzierl:2003ra,Heinrich:2002rc,Anastasiou:2003gr,Gehrmann-DeRidder:2003bm,Binoth:2004jv,Anastasiou:2004qd,Gehrmann-DeRidder:2004tv}
and to  combine them with the recently computed two-loop amplitudes for
parton-parton
scattering~\cite{Bern:2lgggg,Anastasiou:2lqq->qqa,Anastasiou:2lqq->qqb,Anastasiou:2lqg,Glover:2lgg,Anastasiou:2lFBscat,Bern:2lgg->gg,Bern:2lhelsqg,Glover:2lhelsqg,Glover:2lhelsqq},
massless Bhabha scattering~\cite{Bern:2lbhabha}, light-by-light
scattering~\cite{Bern:2001dg,Binoth:2002xg} as well as the gluonic production
of photons~\cite{Bern:2001df} and $\gamma^* \to
3$~partons~\cite{Garland:2lee->3j,Garland:2lhelsee->3j,Moch:2lee->qqg}   to
provide NNLO estimates of the dominant QCD processes in electron-positron
annihilation and hadron-hadron collisions.

One of the recent highlights in perturbative QCD is the succesful computation of
the three-loop contributions to the Altarelli-Parisi kernels that determine the
scale evolution of parton densities.   These splitting functions have recently
been computed in a very impressive series of papers by Moch, Vermaseren and
Vogt~\cite{Moch:2002sn,Moch:nonsinglet,Vogt:singlet} using traditional methods and
superseding previous approximations based on a limited number of 
moments~\cite{vanNeerven:1999ca,vanNeerven:2000uj,vanNeerven:2000wp}. However, an
alternative method using the collinear factorisation properties of QCD
amplitudes has been proposed by Kosower and Uwer~\cite{Kosower:evolker}.   At
this order one encounters contributions from tree amplitudes with four collinear
particles~\cite{DelDuca:treecoll}, one-loop graphs with three collinear
particles~\cite{Catani:triplecoll} and two-loop graphs with two collinear
particle~\cite{Bern:2lsplit}. 

The two-loop splitting functions describing the time-like splitting of one
particle into two collinear massless particles are the subject of this paper.
They have been studied by Bern, Dixon and Kosower~\cite{Bern:2lsplit} who used
the unitarity method of sewing together tree (and one-loop) amplitudes to
construct two-loop amplitudes in the limit where two gluons are collinear both
in the case of QCD and $N=1$ and $N=4$ super Yang-Mills.  In this
approach many of the problems associated with using the light-cone gauge in
traditional Feynman diagram computations of the splitting functions are
avoided. In adition, they have checked that the infrared singularities of
the splitting function agree with the general formula of Catani~\cite{Catani:polestruc,Sterman:multi} and 
provided an ansatz for the form of the non-trivial colour
structures that appear at  ${\cal O}(1/\epsilon)$ therein.

In this paper, we exploit the universal factorisation properties of QCD to
extract the two-loop splitting functions for all parton combinations by taking
the collinear limits of existing two-loop calculations of four-point
helicity
amplitudes~\cite{Berends:hels,DeCausmaecker:hels,Gunion:hels,Xu:hels,Berends:w-vdw} 
and relating them to the two-loop amplitudes for vertex graphs. 
For example, the quark-gluon splitting function is obtained through to  ${\cal
O}(\epsilon^0)$ by taking the collinear limit of the helicity amplitudes for 
$\gamma^* \to q\bar q g$ computed in the `t Hooft-Veltman scheme~\cite{'tHooft:renorm}  and
relating it to the two-loop amplitudes for $\gamma^* \to q\bar q$.  
Similarly, two-loop helicity amplitudes for  $H\to ggg$ and $H \to q\bar q
g$ in th elarge $m_t$ limit~\cite{inprep} give access to the gluon-gluon and quark-antiquark
splitting functions.  The master integrals for these processes involve planar and
non-planar two-loop boxes that have been written in terms of two-dimensional harmonic
polylogarithms~\cite{Gehrmann:3jplanar,Gehrmann:3jnonplanar} (2dHPL) using differential
equations~\cite{Gehrmann:diffeqs}.  In the collinear limit, these 2dHPL collapse
to harmonic polylogarithms of a single variable~\cite{Remiddi:HPL} that can be re-expressed as
Nielsen polylogarithms.

Our paper is organised as follows.   In Section~\ref{sec:notation}, we define
some basic notation for the three- and four-particle processes. Although the
full amplitudes factorise, it is often convenient to work at the level of
colour ordered
amplitudes~\cite{Paton:coldecomp,Cvitanovic:coldecomp,Berends:w-vdw,Mangano:gluoncol,Bern:coldecomp,DelDuca:coldecomp}.   
For the processes at hand, the colour structure is
particularly simple and the colour ordered amplitudes are straightforwardly
obtained by stripping away the colour matrices.  The collinear singularities of
the full amplitude are then obtained by restoring these colour matrices. The
collinear limit of the colour ordered amplitudes is described in Section~\ref{sec:collinear}
where we define the splitting functions in terms of the momenta and helicities
of the collinear particles and also specify how the collinear limit is
approached. The procedure for renormalising the splitting functions is also
described.  We turn to the quark-gluon collinear limit in
Section~\ref{sec:quarkgluon} and treat the gluon-gluon and quark-antiquark
collinear limits in Sections~\ref{sec:gluongluon} and
\ref{sec:quarkantiquark}.  In each case we show how the general helicity
structure of the four-particle amplitude behaves in the singular limit and give results for
the tree, one-loop and two-loop splitting functions.  In the case of
gluon-gluon splitting, we recover the QCD results of Ref.~\cite{Bern:2lsplit}
in the 't Hooft-Veltman scheme where external particles are treated in
4-dimensions while the particles internal to the loop are in
$(4-2\epsilon)$-dimensions.  We examine the limits as $w \to 0$ and $w\to 1$ in
Sec.~\ref{sec:limits} and show that the soft behaviour of the two-loop
splitting functions is well behaved and  consistant with the expected
behaviour. As a by product of our calculation, we provide an expression for the
two-loop soft splitting function in Section~\ref{sec:soft}. Finally, our main
results are summarized in Section~\ref{sec:conclusions} and accompanied by
concluding remarks.  The one-loop singularity operators, ${\bom I}_1$ in Catani's language,
for each of the processes studied here are detailed in the appendix.

\section{Notation}
\label{sec:notation}

We consider the soft and collinear limits of two-loop matrix elements 
by taking the limits of known matrix elements.
In general, we consider the decay of a massive object $X$ into three partons,
\begin{equation}
X (q) \longrightarrow a(p_a) + b (p_b) + c(p_c),
\end{equation}
where $X$ is a virtual photon or Higgs boson and $a$, $b$ and  $c$ are massless partons.
Specifically, we have 
\begin{eqnarray}
\label{eq:gamma3}
\gamma^*  &\longrightarrow& q  + \bar q   + g ,\\
\label{eq:Hggg}
H &\longrightarrow& g  + g   + g ,\\
\label{eq:Hqqg}
H &\longrightarrow& q    + g + \bar q .
\end{eqnarray}
In the collinear limits, the matrix elements factorise onto the two particle final states,
\begin{equation}
X (q) \longrightarrow A(p_A) + B (p_B),
\end{equation}
or equivalently
\begin{eqnarray}
\label{eq:gamma2}
\gamma^*   &\longrightarrow& Q  + \bar Q,\\
\label{eq:Hgg}
H &\longrightarrow& G + G  .
\end{eqnarray}
The kinematics of these processes is fully described by the invariants
\begin{equation}
\sab = (p_a+p_b)^2\;, \qquad \sac = (p_a+p_c)^2\;, \qquad 
\sbc = (p_b+p_c)^2\;,
\end{equation}
and
\begin{equation}
\sabc = (p_a+p_b+p_c)^2 = (p_A+p_B)^2 = \sAB,
\end{equation}
which fulfil
\begin{equation}
\sab + \sac + \sbc \equiv \sabc \equiv \sAB.
\end{equation}

The {\em unrenormalised}
amplitude for process ${\cal P}$ is a vector in colour space and can be written
in terms of a colour-stripped amplitude,
\begin{equation}
|{\cal M}_{\P}\rangle = {\bom C_{\P}} {\cal A}_{\P}.
\end{equation}
For the processes at hand the overall colour structures are given by,
\begin{eqnarray}
{\bom C_{\gamma^* \to Q\bar Q}} &=&   ~{\bom \delta_{AB}},\\
{\bom C_{\gamma^* \to q\bar q g}} &=&   ~{\bom T^c_{ab}},\\
{\bom C_{H \to G  G}} &=& ~Tr(\bom T^{A}}{\bom T^{B}),\\
{\bom C_{H \to g g g}} &=&  ~{\bom f^{abc}},\\
{\bom C_{H \to q \bar q g  }} &=& ~{\bom T^c_{ab}},
\end{eqnarray}
where upper and lower indices represent colour 
indices in the adjoint and fundamental 
representations respectively.
Note that the colour ordering of the two- and three-particle amplitudes is
trivial here.  
The normalisation $Tr(\bom T^A\bom T^B)={\bom \delta^{AB}}/{2}$ is 
used throughout. 
While the explicit calculations presented here will be derived from the
processes of Eqs.~(\ref{eq:gamma3}) --(\ref{eq:Hgg}), the universal
factorisation properties of QCD amplitudes ensure that the singular limits will
be applicable to
processes with more particles and more complicated colour structures.

The colour ordered amplitudes have a perturbative
expansion in terms of the colour stripped $i$-loop amplitude
$|{\cal A}_{\P}^{(i)}\rangle$,
\begin{equation}
{\cal A}_{\P} = C_{\P} \left[
{\cal A}_{\P}^{(0)} 
+ \left(\frac{\alpha_s}{2\pi}\right) {\cal A}_{\P}^{(1)} 
+ \left(\frac{\alpha_s}{2\pi}\right)^2 {\cal A}_{\P}^{(2)} 
+ {\cal O}(\alpha_s^3) \right] \;,
\label{eq:renorme}
\end{equation}
where $C_{\P}$ contains the overall couplings such that
\begin{eqnarray}
C_{\gamma^* \to Q\bar Q} &=& \sqrt{4\pi\alpha} ,\\
C_{\gamma^* \to q\bar q g} &=& \sqrt{4\pi\alpha}\sqrt{4\pi\alpha_s} ,\\
C_{H \to G  G} &=& C_1,\\
C_{H \to g g g} &=& C_1\sqrt{4\pi\alpha_s} ,\\
C_{H \to q  \bar q g  } &=& C_1\sqrt{4\pi\alpha_s} ,
\end{eqnarray}
and where $C_1$ is the effective Higgs-gluon-gluon coupling.

\section{The collinear limit}
\label{sec:collinear}

When two particles become collinear, colour ordered amplitudes factorise
in a universal way~\cite{Bern:split1e0,Bern:split1gluon,Bern:split1QCD}.   
If we focus on particles $a$ and $b$ with helicities 
$\lambda_a$ and $\lambda_b$ becoming
collinear so that $\sab \to 0$ and forming particle $A$ with helicity $\lambda_A$, then at tree-level we find
that,
\begin{eqnarray}
\label{eq:treecol}
{\cal A}^{(0)}(a^{\lambda_a},b^{\lambda_b},\ldots) &\stackrel{a||b}{\longrightarrow}& 
\phantom{+}\sum_{\lambda_A=\pm}\Split^{A \to ab,(0)}({-\lambda_A},a^{\lambda_a},b^{\lambda_b}) 
{\cal A}^{(0)}(A^{\lambda_A},\ldots).
\end{eqnarray}

\FIGURE[t!]{
\label{fig:tree}
\epsfig{file=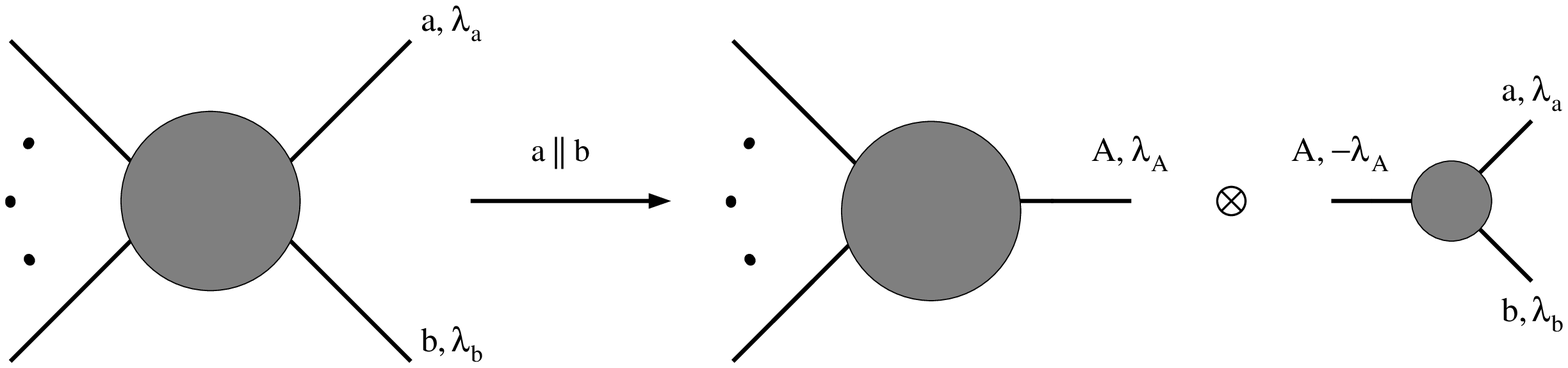,width=12cm}
\caption{The collinear behaviour of a tree-level amplitude.  Particles $a$ and $b$ form (a slightly offshell)
particle $A$.  The singular behaviour is encapsulated in the right hand vertex and is denoted by
$\Split^{A \to ab,(0)}({-\lambda_A},a^{\lambda_a},b^{\lambda_b})$.
}
}
The factorisation of the $(n+1)$-particle colour ordered
tree amplitude into a $n$-particle tree amplitude multiplied by the splitting function 
is illustrated in Fig.~\ref{fig:tree}.  The singular factors are all contained in 
$\Split^{A \to ab,(0)}({-\lambda_A},a^{\lambda_a},b^{\lambda_b})$ which behaves as $1/\sqrt{s_{ab}}$.   
The momentum of the slightly off-shell particle $A$ links the two factors, $$p_A =p_a+p_b.$$
while the helicity $\lambda_A$ is summed over. However, for some helicity combinations, such as the case of a 
positive helicity gluon splitting into two negative helicity gluons,  the tree splitting function vanishes.

Similarly, at one- and two-loops, the colour stripped amplitudes factorise as follows,
\begin{eqnarray}
\label{eq:onecol}
{\cal A}^{(1)}(a^{\lambda_a},b^{\lambda_b},\ldots) &\stackrel{a||b}{\longrightarrow}& 
\phantom{+}\sum_{\lambda_A=\pm}\Split^{A \to ab,(1)}({-\lambda_A},a^{\lambda_a},b^{\lambda_b})
{\cal A}^{(0)}(A^{\lambda_A},\ldots)\nonumber\\
&&+\sum_{\lambda_A=\pm}\Split^{A \to ab,(0)}({-\lambda_A},a^{\lambda_a},b^{\lambda_b}) 
{\cal A}^{(1)}(A^{\lambda_A},\ldots),\\
{\cal A}^{(2)}(a^{\lambda_a},b^{\lambda_b},\ldots) &\stackrel{a||b}{\longrightarrow}& 
\label{eq:twocol}
\phantom{+}\sum_{\lambda_A=\pm}\Split^{A \to ab,(2)}({-\lambda_A},a^{\lambda_a},b^{\lambda_b}) 
{\cal A}^{(0)}(A^{\lambda_A},\ldots)\nonumber\\
&&+\sum_{\lambda_A=\pm}\Split^{A \to ab,(1)}({-\lambda_A},a^{\lambda_a},b^{\lambda_b})
{\cal A}^{(1)}(A^{\lambda_A},\ldots)\nonumber \\
&&+\sum_{\lambda_A=\pm}\Split^{A \to ab,(0)}({-\lambda_A},a^{\lambda_a},b^{\lambda_b})
{\cal A}^{(2)}(A^{\lambda_A},\ldots),
\end{eqnarray}
where $\Split^{A \to ab,(i)}({-\lambda_A},a^{\lambda_a},b^{\lambda_b})$ denotes the $i$-loop 
splitting function.  These factorisations are illustrated in Figs.~\ref{fig:one} and \ref{fig:two}.

In addition to the explicit helicity and particle-type dependence, 
all of the splitting functions also depend on 
the momentum fraction $w$ carried by each particle,
\begin{eqnarray}
\label{eq:colmom}
p_a &\to& (1-w) p_A+k_\perp-\frac{k_\perp^2}{1-w}\frac{n}{ 2p_A\cdot n}, \nonumber \\
p_b &\to& w p_A-k_\perp-\frac{k_\perp^2}{w}\frac{n}{ 2p_A\cdot n}, \nonumber \\
s_{ab} &=& -\frac{k_\perp^2}{w(1-w)}\to 0,
\end{eqnarray}
where $k_\perp$ is the transverse momentum and 
$n$ is an arbitrary lightlike momentum such that
$2Q\cdot k_\perp = 2 n\cdot k_\perp = 0$,
and implicitly on the dimensional regularisation parameter $\epsilon$.

\FIGURE[t!]{
\label{fig:one}
\epsfig{file=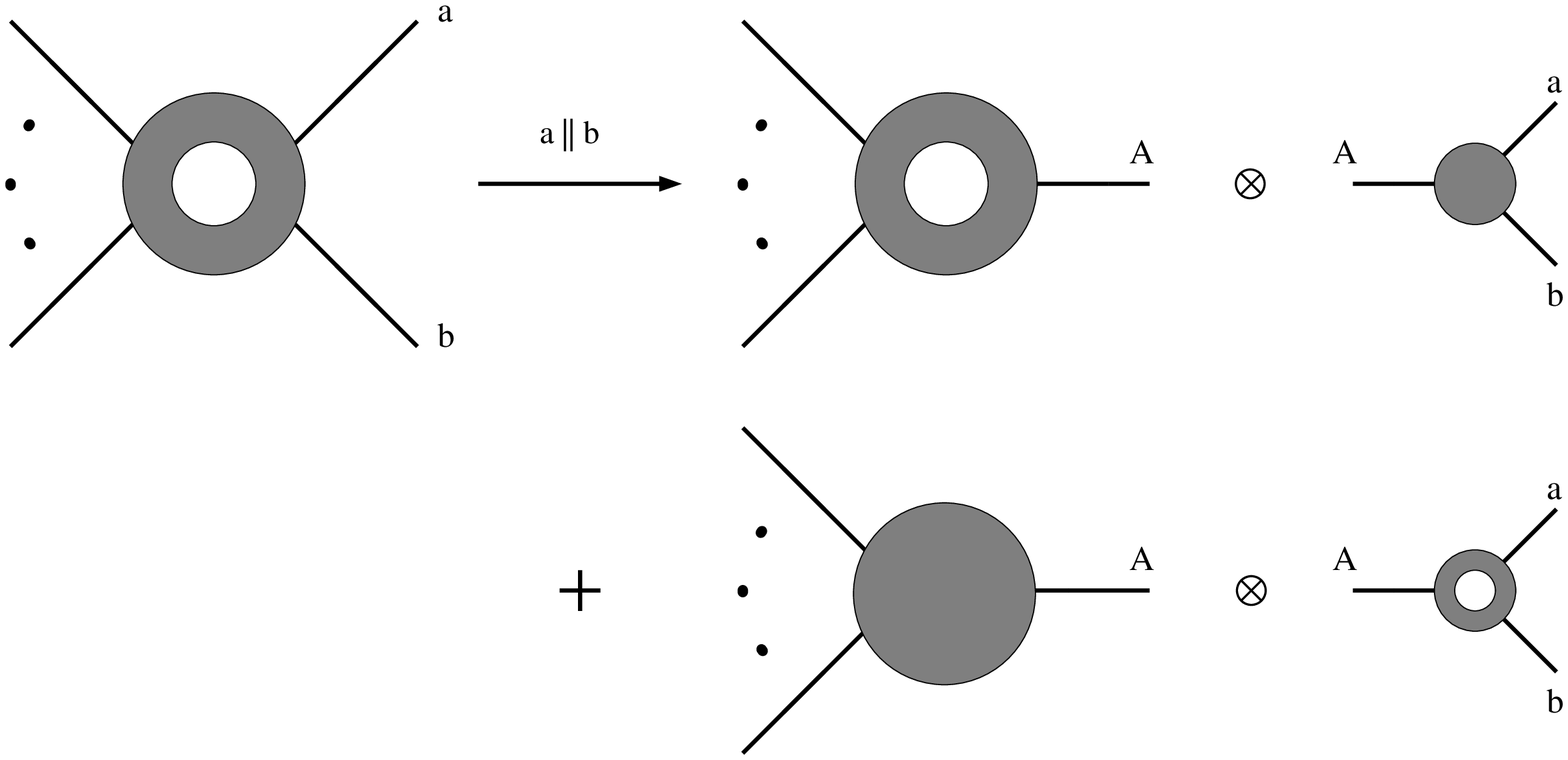,width=12cm}\\
\caption{The collinear behaviour of a one-loop amplitude.
}
}
\FIGURE[t!]{
\label{fig:two}
\epsfig{file=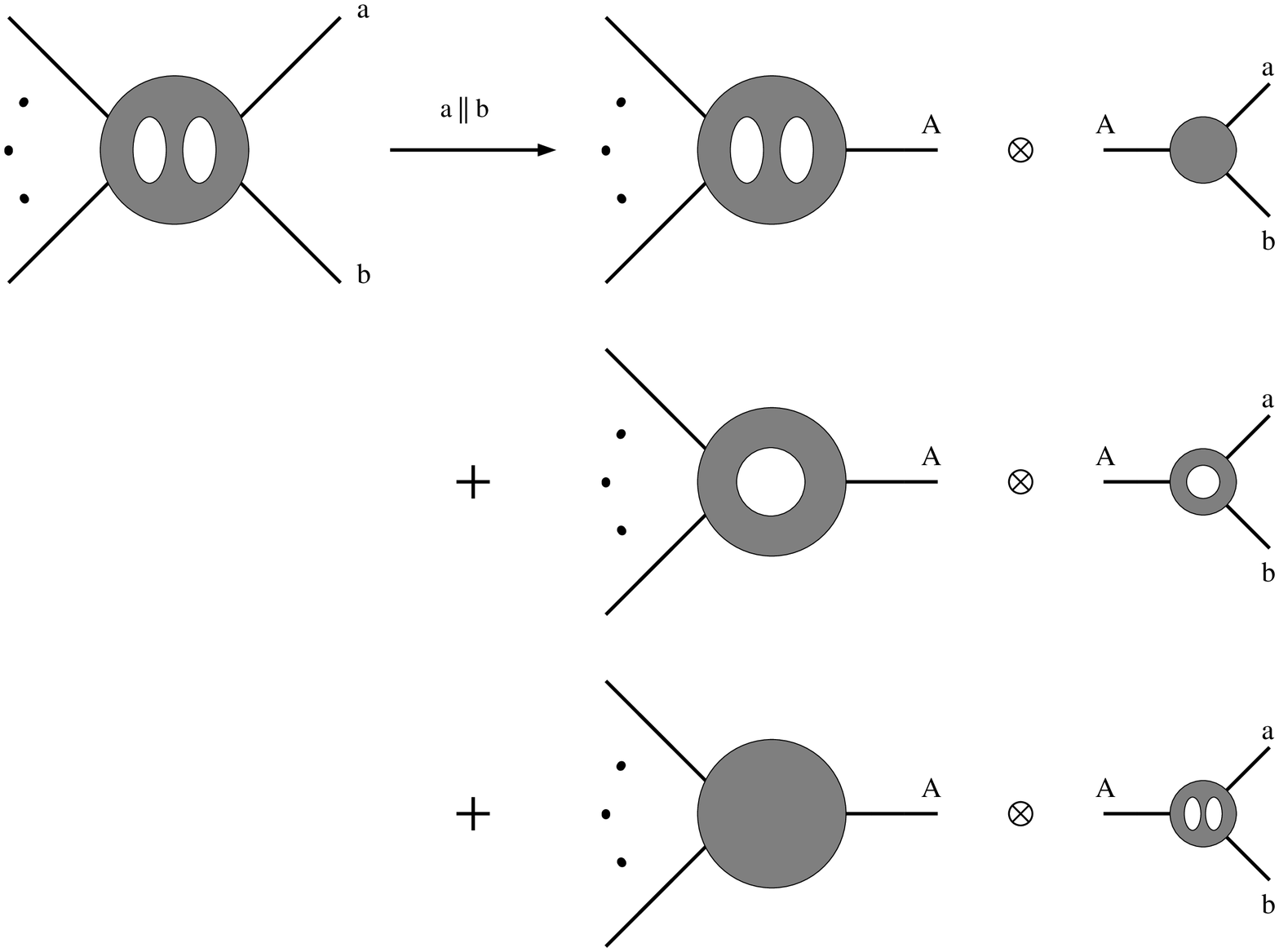,width=12cm}\\
\caption{The collinear behaviour of a two-loop amplitude.
}
}

The tree and one-loop splitting functions have been known for some time~\cite{Mangano:multipart,Bern:split1e0,Bern:split1gluon,Bern:split1QCD}, however the form of the 
two-loop splitting function is only known in the case where a gluon split into gluons~\cite{Bern:2lsplit}. 

When the tree splitting function does not vanish, it is customary to factor it out,  
\begin{eqnarray}
\label{eq:r1def}
\Split^{A \to ab,(1)}({-\lambda_A},a^{\lambda_a},b^{\lambda_b}) 
&=& r_{S}^{A \to ab,(1)}({-\lambda_A},a^{\lambda_a},b^{\lambda_b})
\Split^{A \to ab,(0)}({-\lambda_A},a^{\lambda_a},b^{\lambda_b}),
\end{eqnarray}
such that all of the infrared singularities lie in the ratio factor $r_{S}^{A \to ab,(1)}$.
This factor contains all of the non-trivial $w$ dependence as well as the scale violating 
factor, $$\left(\frac{4\pi\mu^2}{s_{ab}}\right)^{\e}$$ that modifies the $1/\sqrt{s_{ab}}$ behaviour of the
tree-level splitting function.   $r_{S}^{A \to ab,(1)}$ also contains the infrared poles produced by soft and
collinear particles circulating in the loop.  Explicit formulae, valid to all orders in the dimensional
regularisation parameter are given in Refs.~\cite{Bern:split1gluon,Bern:split1QCD}.

Similarly, at two-loops,  when the tree-level splitting function does not vanish,
the singularities are factored into $r_{S}^{A \to ab,(2)}$,
\begin{eqnarray}
\label{eq:r2def}
\Split^{A \to ab,(2)}({-\lambda_A},a^{\lambda_a},b^{\lambda_b}) 
&=& r_{S}^{A \to ab,(2)}({-\lambda_A},a^{\lambda_a},b^{\lambda_b})
\Split^{A \to ab,(0)}({-\lambda_A},a^{\lambda_a},b^{\lambda_b}).
\end{eqnarray}
Catani~\cite{Catani:polestruc} has shown how to organize the 
infrared pole structure of the two-loop contributions renormalized in the 
\MSbar\ scheme in terms of the tree and renormalized one-loop amplitudes.
Motivated by the structure of Catani's formula and by the results for planar amplitudes in maximally supersymmetric
Yang-Mills theory~\cite{Anastasiou:planar}, 
we write the singular behaviour of the {\em unrenormalised} $r_{S}^{A \to ab,(2)}$ as (dropping the helicity labels),
\begin{eqnarray}
\label{eq:r2}
r_{S}^{A \to ab,(2)}(\e) &=& 
\frac{1}{2}\left(r_{S}^{A \to ab,(1)}(\e)\right)^2
+\frac{e^{-\epsilon\gamma}c_{\Gamma}(\e)}{c_{\Gamma}(2\e)}\left(\frac{\beta_0}{\e}+K\right )r_{S}^{A \to ab,(1)}(2\e)
+\Delta H^{A \to ab,(2)}(\e)\nonumber \\
&&
+r_{S}^{A \to ab,(2),fin}(\epsilon) + {\cal O}(\epsilon),
\end{eqnarray}
where the Euler constant$\gamma = 0.5772\ldots$ and
\begin{equation}
c_{\Gamma}(\e) = \frac{\Gamma(1-\e)^2\Gamma(1+\e)}{\Gamma(1-2\e)}.
\end{equation}
The first two terms contain all infrared singularities from ${\cal O}(1/\e^4)$ through to ${\cal O}(1/\e^2)$
while the remaining single poles are absorbed in $\Delta H^{A \to ab,(2)}(\e)$,
\begin{equation}
\Delta H^{A \to ab,(2)}(\e) = \frac{e^{-\epsilon\gamma}c_\Gamma(\e) }{4 \epsilon}
\left(\frac{4\pi\mu^2}{-s_{ab}}\right)^{2\epsilon}\big (w(1-w)\big)^{-2\e}
\left(H^{(2)}_a+H^{(2)}_b-H^{(2)}_A -\beta_0 K +\beta_1\right).
\end{equation}
For splitting functions involving only two collinear particles, 
there are no colour correlations in
$\Delta H^{A \to ab,(2)}(\e)$.
Here, $H_q$ ($H_{\bar q}$) and $H_g$ are the usual constants appearing in Catani's {\em renormalised} formula,
\begin{eqnarray}
\label{eq:defHq}
H_{q}^{(2)} =H_{\bar q}^{(2)} &=&\left(\frac{\pi^2}{2}-6 ~\zeta_3 
-\frac{3}{8}\right) \CF^2
+\left(\frac{13}{2}\zeta_3 +\frac{245}{216}-\frac{23}{48} \pi^2 \right) \CA \CF
\nonumber \\
&& + \left(-\frac{25}{54}+\frac{\pi^2}{12} \right) T_R \NF \CF, \\
\label{eq:defHg}
H_{g}^{(2)} &=& 
\frac{20}{27} T_R^2 \NF^2
+ T_R \CF \NF
-\left(\frac{ \pi^2}{36}+\frac{58}{27} \right)T_R \NF\CA
\nonumber \\ &&
+\left(\frac{\zeta_3}{2}+\frac{5}{12}
+\frac{11}{144}\pi^2 \right) \CA^2,
\end{eqnarray}
while the constants $\beta_0$, $\beta_1$ and $K$ are
\begin{eqnarray}
\beta_0 &=& \frac{11\CA-2N_F}{6},\\
\beta_1 &=& \frac{17\CA^2-10\CA T_R\NF-6\CF T_R\NF}{6},\\
K &=& \left( \frac{67}{18} - \frac{\pi^2}{6} \right) \CA -  \frac{10}{9} T_R \NF.
\end{eqnarray}
For SU(N) gauge theory, $\CA = N$, $\CF = (N^2-1)/2/N$ and $T_R = 1/2$.
Note that while the pole contribution to $\Delta H^{A \to ab,(2)} = \left(H^{(2)}_a+H^{(2)}_b-H^{(2)}_A -\beta_0 K
+\beta_1\right)/\e$ is fixed,
there is considerable freedom in defining the ${\cal O}(\e) $ contribution to 
the factor multiplying the pole.
.  Different choices affect the finite remainder.
We choose this particular factor since it regulates the
$s_{ab} \to 0$, $w \to 0$ and $w \to 1$ limits of the splitting function.

The remaining finite contributions are contained in $r_{S}^{A \to ab,(2),fin}(\e)$. As we will demonstrate,
explicit calculation shows that Eq.~(\ref{eq:r2}) fixes the infrared pole structure of the splitting
function exactly, leaving only $r_{S}^{A \to ab,(2),fin}(\e)$ to be determined.

\subsection{Ultraviolet renormalization}
\label{subsec:renorm}

The splitting functions have the perturbative expansion,
\begin{equation}
\Split^{A \to ab}  = \sqrt{4\pi\alpha_s}\left[
\Split^{A \to ab,(0)}  
+ \left(\frac{\alpha_s}{2\pi}\right) \Split^{A \to ab,(1)}  
+ \left(\frac{\alpha_s}{2\pi}\right)^2 \Split^{A \to ab,(2)} 
+ {\cal O}(\alpha_s^3) \right],
\end{equation}

The renormalization of the splitting function is carried out by replacing 
the bare coupling $\alpha_s$ with the renormalized coupling 
$\alpha_s(\mu^2)$,
evaluated at the renormalization scale $\mu^2$
\begin{equation}
\alpha_s\mu_0^{2\e} S_\e = \alpha_s(\mu^2) \mu^{2\e}\left[
1- \frac{\beta_0}{\e}\left(\frac{\alpha_s(\mu^2)}{2\pi}\right) 
+\left(\frac{\beta_0^2}{\e^2}-\frac{\beta_1}{2\e}\right)
\left(\frac{\alpha_s(\mu^2)}{2\pi}\right)^2+{\cal O}(\alpha_s^3) \right]\; ,
\end{equation}
where
\begin{displaymath}
S_\e =(4\pi)^\e e^{-\e\gamma},
\end{displaymath}
and $\mu_0^2$ is the mass parameter introduced 
in dimensional regularization~\cite{dreg1,dreg2,'tHooft:renorm} to maintain a 
dimensionless coupling 
in the bare QCD Lagrangian density.

Applying the renormalisation procedure, we have 
\begin{eqnarray}
\Split^{A \to ab,(0),R}  &=& \Split^{A \to ab,(0)} ,
 \nonumber \\
\Split^{A \to ab,(1),R}  &=& 
S_\e^{-1} \Split^{A \to ab,(1)} 
-\frac{\beta_0}{2\e} \Split^{A \to ab,(0)}  ,  \nonumber \\
\Split^{A \to ab,(2),R} &=& 
S_\e^{-2} \Split^{A \to ab,(2)}  
-\frac{3\beta_0}{2\e} S_\e^{-1}
\Split^{A \to ab,(1)}  
-\left(\frac{\beta_1}{4\e}-\frac{3\beta_0^2}{8\e^2}\right)
\Split^{A \to ab,(0)}.
\end{eqnarray}
Equivalently, for the amplitudes where the tree splitting function does not
vanish,
\begin{eqnarray}
r_S^{A \to ab,(1),R}  &=& 
S_\e^{-1} r_S^{A \to ab,(1)} 
-\frac{\beta_0}{2\e}    ,  \nonumber \\
r_S^{A \to ab,(2),R} &=& 
S_\e^{-2} r_S^{A \to ab,(2)}  
-\frac{3\beta_0}{2\e} S_\e^{-1}
r_S^{A \to ab,(1)}  
-\left(\frac{\beta_1}{4\e}-\frac{3\beta_0^2}{8\e^2}\right).
\end{eqnarray}
Applying this transformation to the unrenormalised splitting ratios, we find
that,
\begin{eqnarray}
\label{eq:r2ren}
r_{S}^{A \to ab,(2),R}(\e) &=& 
\frac{1}{2}\left(r_{S}^{A \to ab,(1),R}(\e)\right)^2
-\frac{\beta_0}{\e}r_{S}^{A \to ab,(1),R}(\e)
+\frac{e^{-\epsilon\gamma}c_{\Gamma}(\e)}{c_{\Gamma}(2\e)}\left(\frac{\beta_0}{\e}+K\right )r_{S}^{A \to ab,(1),R}(2\e)\nonumber \\
&&
+\Delta H^{A \to ab,(2),R}(\e) 
+r_{S}^{A \to ab,(2),R,fin}(\epsilon) + {\cal O}(\epsilon),
\end{eqnarray}
where,
\begin{equation}
\Delta H^{A \to ab,(2),R}(\e) = \frac{e^{-\epsilon\gamma}c_\Gamma(\e) }{4 \epsilon}
\left(\frac{4\pi\mu^2}{-s_{ab}}\right)^{2\epsilon}\big (w(1-w)\big)^{-2\e}
\left(H^{(2)}_a+H^{(2)}_b-H^{(2)}_A\right).
\end{equation}
together with a modification of the finite remainder, $r_{S}^{A \to
ab,(2),R,fin}(\epsilon)$.
Up to trivial gamma function coefficients, which only modify the finite
remainder, Eq.~(\ref{eq:r2ren}) has the same structure as that given for the
renormalised gluon splitting function in  Ref.~\cite{Bern:2lsplit} and also
Catani's formula for the 
general structure of renormalised two-loop amplitudes~\cite{Catani:polestruc}. 
Throughout the remainder of the paper we will work with unrenormalised amplitudes.

\section{The quark-gluon splitting amplitudes}
\label{sec:quarkgluon}

The unrenormalized 
amplitude $|{\cal M}\rangle$ for the virtual 
photon initiated processes, (\ref{eq:gamma2}) and 
(\ref{eq:gamma3}),
can be written as
\begin{eqnarray}
\label{eq:Mdef}
|{\cal M}_{\gamma^*\to Q\bar Q}\rangle &=& {\bom \delta_{AB}}\, V^\mu \S_\mu(Q;\bar Q),\nonumber \\
|{\cal M}_{\gamma^*\to q\bar q g}\rangle &=& {\bom T^c_{ab}}\,V^\mu \S_\mu(q;g;\bar q)\; ,
\end{eqnarray}
where $V^\mu$ represents the lepton current  and $\S_\mu$ denotes the colour
ordered hadron current.
Using the Weyl-van der Waerden spinor notation (see Ref.~\cite{Berends:w-vdw}), 
the hadronic current $\S_{\mu}$ is related to the fixed helicity 
currents, $\S_{{\dot A}B}$, by
\begin{eqnarray}
\label{eq:scurqq}
\S_{\mu}(Q+;\overline{Q}-) &=& R^V_{f_1f_2}
\, \sigma_{\mu}^{{\dot A}B} \S_{{\dot A}B}(Q+;\overline{Q}-) ,\\
\S_{\mu}(Q-;\overline{Q}+) &=& L^V_{f_1f_2}
\, \sigma_{\mu}^{{\dot A}B} \S_{{\dot A}B}(Q-;\overline{Q}+) ,\\
\label{eq:scurqqg}
\S_{\mu}(q+;g\lambda;\overline{q}-) &=& R^V_{f_1f_2}
 \sqrt2\, \sigma_{\mu}^{{\dot A}B} \S_{{\dot A}B}(q+;g\lambda;\overline{q}-) ,\\
 \S_{\mu}(q-;g\lambda;\overline{q}+) &=& L^V_{f_1f_2}
 \sqrt2\, \sigma_{\mu}^{{\dot A}B} \S_{{\dot A}B}(q-;g\lambda;\overline{q}+) .
\end{eqnarray}

Labelling the particle momenta by their type, and dropping terms that vanish
when contracted with the lepton current, we find that
\begin{eqnarray}
\label{eq:helampzqq}
{\S}_{\dot AB}(Q+;\overline{Q}-)&=& A~Q_{\dot A}\overline{Q}_{B} \\
\label{eq:helampzqqg1}
{\S}_{\dot AB}(q+;g+;\overline{q}-)
&=&
\alpha(y,z)~
\frac{q_{\dot AD}\bar{q}^D \bar{q}_{B}}{\langle qg \rangle\langle g\bar{q} \rangle}
+\beta(y,z)~
\frac{g_{\dot AD} \bar{q}^D \bar{q}_{B}}{\langle qg \rangle\langle g\bar{q} \rangle}
+\gamma(y,z)~
\frac{q_{\dot CB}g^{\dot C}g_{\dot A}}{\langle qg \rangle\langle g\bar{q} \rangle^*}  
\;,  \\
\label{eq:helampzqqg2}
{\S}_{\dot AB}(q+;g-;\overline{q}-)
&=&
-\alpha(z,y)~
\frac{\bar{q}_{\dot DB}q^{\dot D} q_{\dot A}}{\langle qg \rangle^*\langle g\bar{q} \rangle^*}
-\beta(z,y)~
\frac{g_{\dot DB} q^{\dot D} q_{\dot A}}{\langle qg \rangle^*\langle g\bar{q} \rangle^*}
-\gamma(z,y)~
\frac{\bar{q}_{\dot AC}g^{C}g_{B}}{\langle \bar{q}g \rangle^*\langle gq \rangle}  
\;.
\end{eqnarray}
In these expressions, $\langle ij\rangle = \langle i-| j+\rangle$ and 
$[ ij] = \langle i+| j-\rangle$ where $|i \lambda\rangle$ is the Weyl spinor for a massless particle with 
momentum $i$ and helicity $\lambda$.  The spinor products are antisymmetric and satisfy
$[ ij] = - \langle ij\rangle^*$ and $\langle ij\rangle[ji] = s_{ij}$.

The currents with the quark helicities flipped
follow from parity conservation:
\begin{eqnarray}
\label{eq:quarkflip}
\S_{{\dot A}B}(Q-;\overline{Q}+) &=&
( \S_{{\dot B}A}(Q+;\overline{Q}-))^*,\nonumber \\
\S_{{\dot A}B}(q-;g\lambda;\overline{q}+) &=&
( \S_{{\dot B}A}(q+;g(-\lambda);\overline{q}-))^*.
\end{eqnarray}

The coefficients $A$ and $\Omega$ ($\Omega = \alpha,\beta,\gamma$) contain 
all of the integrals over loop
momenta and ultimately determine the infrared structure of the amplitude.
$A$ depends only on the invariant mass of the virtual photon $s_{AB}$, 
while  $\Omega$ depends on $y=s_{qg}/s_{qg\bar q}$ and $z=s_{g\bar q}/s_{qg\bar q}$.    
Each coefficient is a linear combination of master
loop integrals~\cite{Gehrmann:3jplanar,Gehrmann:3jnonplanar} with coefficients that are rational functions of the 
scale invariants and the spacetime dimension.   
Expansions of the renormalised $\Omega$ around $\e=0$ are given in Ref.~\cite{Garland:2lee->3j,Garland:2lhelsee->3j,Moch:2lee->qqg}.

Taking the quark-gluon collinear limit corresponds to $y = s_{qg}/s_{qg\bar q} \to 0$,
with the quark and gluon carrying momentum fractions $(1-w)Q$ and 
$w Q$ respectively. When the quark and gluon both have positive helicity we find that,
\begin{eqnarray}
\label{eq:zqqglim1}
{\S}_{\dot AB}(q+;g+;\overline{q}-)
&\stackrel{q||g}{\longrightarrow}&
\frac{1}{\sqrt{w}~\langle qg \rangle}
\biggl[
\alpha(y,z)\Big|_{y \to 0}~(1-w)
+\beta(y,z)\Big|_{y \to 0}~w
\biggr]~Q_{\dot A}\overline{Q}_B.
\end{eqnarray}
Similarly, when the quark has positive helicity but the gluon has negative helicity, we find,  
\begin{eqnarray}
\label{eq:zqqglim2}
{\S}_{\dot AB}(q+;g-;\overline{q}-)
&\stackrel{q||g}{\longrightarrow}&
-\frac{1-w}{\sqrt{w}~[qg]}\biggl[
\alpha(z,y)\Big|_{y \to 0}  
\biggr]~Q_{\dot A}\overline{Q}_B. 
\end{eqnarray}
We see that in both cases we recover the two-particle helicity structure of Eq.~(\ref{eq:zqqglim1}).

In the antiquark-gluon collinear limit,  $z = s_{g\bar q}/s_{qg\bar q} \to 0$ and
the antiquark and gluon carrying momentum fractions $(1-w)Q$ and 
$w Q$ respectively.
When the antiquark has negative helicity then, 
\begin{eqnarray}
\label{eq:zqqglim2a}
{\S}_{\dot AB}(q+;g-;\overline{q}-)
&\stackrel{\bar{q}||g}{\longrightarrow}&
-\frac{1}{\sqrt{w}~[ g\bar q ]}
\biggl[
\alpha(z,y)\Big|_{z \to 0}~(1-w)
+\beta(z,y)\Big|_{z \to 0}~w
\biggr]~Q_{\dot A}\overline{Q}_B,\\
\label{eq:zqqglim2b}
{\S}_{\dot AB}(q+;g+;\overline{q}-)
&\stackrel{\bar{q}||g}{\longrightarrow}&
\frac{1-w}{\sqrt{w}~\langle g \bar q\rangle}\biggl[
\alpha(y,z)\Big|_{z \to 0}  
\biggr]~Q_{\dot A}\overline{Q}_B. 
\end{eqnarray}

The helicity amplitude coefficients have the perturbative expansions,
\begin{equation}
A =  C_{\gamma^* \to Q\bar Q}\, \left[
A^{(0)}  
+ \left(\frac{\alpha_s}{2\pi}\right) A^{(1)}  
+ \left(\frac{\alpha_s}{2\pi}\right)^2 A^{(2)} 
+ {\cal O}(\alpha_s^3) \right],
\end{equation}
and
\begin{equation}
\Omega(y,z) =  C_{\gamma^* \to q\bar q g}\, \left[
\Omega^{(0)}(y,z)  
+ \left(\frac{\alpha_s}{2\pi}\right) \Omega^{(1)}(y,z)  
+ \left(\frac{\alpha_s}{2\pi}\right)^2 \Omega^{(2)}(y,z) 
+ {\cal O}(\alpha_s^3) \right] \;,\nonumber \\
\end{equation}
for $\Omega = \alpha,\beta,\gamma$.

\subsection{The tree-level quark-gluon splitting amplitudes}

At leading order,
\begin{equation}
\alpha^{(0)}(y,z) = \beta^{(0)}(y,z) = 1\qquad
\mbox{and}\qquad A^{(0)}=1.
\end{equation}
Bearing in mind that the splitting function defined in Eq.~(\ref{eq:treecol}) relates the colour ordered
amplitudes,  comparing Eqs.~(\ref{eq:zqqglim1}) and (\ref{eq:zqqglim2}) with Eq.~(\ref{eq:helampzqq}),
we can immediately read off the tree-level quark-gluon splitting functions for colour stripped amplitudes,
\begin{eqnarray}
\label{eq:qgtree}
\Split^{Q \to qg,(0)}(-,q^{+},g^{+})&=&
\frac{\sqrt{2}}{\sqrt{w} ~\langle qg \rangle},\nonumber \\
\Split^{Q \to qg,(0)}(-,q^{+},g^{-})&=&
\frac{-\sqrt{2}\, (1-w)}{\sqrt{w} ~[ qg ]}.
\end{eqnarray}
The factors of $\sqrt{2}$ are remnants of the definition of the helicity current in Eq.~(\ref{eq:scurqqg}).
In addition, the soft gluon singularity as $w \to 0$ is explicit.

Splitting functions where $Q$ has negative helicity are obtained by complex
conjugation according to Eq.~(\ref{eq:quarkflip}),
\begin{eqnarray}
\label{eq:qgtreeflip}
\Split^{Q \to qg,(0)}(+,q^{-},g^{-})&=&
\frac{-\sqrt{2}}{\sqrt{w} ~[ qg ]},\nonumber \\
\Split^{Q \to qg,(0)}(+,q^{-},g^{+})&=&
\frac{\sqrt{2}\, (1-w)}{\sqrt{w} ~\langle qg \rangle}.
\end{eqnarray}

Similarly, the tree splitting functions for an antiquark $\bar Q$ to split into a
gluon $g$ and antiquark $\bar q$ are given by,
\begin{eqnarray}
\Split^{\bar Q \to g\bar q,(0)}(+,g^{+},\bar{q}^{-})=
\phantom{-}\frac{\sqrt{2}}{\sqrt{w}~[ g\bar q ]},
&&
\Split^{\bar Q \to g\bar q,(0)}(+,g^{-},\bar{q}^{-})=
-\frac{\sqrt{2}(1-w)}{\sqrt{w}~\langle g \bar q\rangle} 
\\
\Split^{\bar Q \to g\bar q,(0)}(-,g^{-},\bar{q}^{+})=
-\frac{\sqrt{2}}{\sqrt{w}~\langle g\bar q \rangle},
&&\Split^{\bar Q \to g\bar q,(0)}(-,g^{+},\bar{q}^{+})=
\phantom{-}\frac{\sqrt{2}(1-w)}{\sqrt{w}~[ g \bar q]}.
\end{eqnarray}

\subsection{The one-loop quark-gluon splitting amplitudes}

The unrenormalised one-loop splitting functions defined in Eq.~(\ref{eq:r1def}) 
are presented in Ref.~\cite{Bern:split1QCD} in  a variety of schemes
to all orders in $\e$, and we reproduce them here in the 't Hooft-Veltman regularisation scheme
($\delta_R = 1$) for the sake of completeness,
\begin{eqnarray}
\label{eq:qgone}
\lefteqn{r_{S}^{Q \to qg,(1)}(\mp,q^{\pm},g^{\pm})
=r_{S}^{\overline{Q} \to g\bar q,(1)}(\mp,g^{\pm},\bar{q}^{\pm})=}\nonumber \\
&&
-c_{\Gamma}(\e)\left(\frac{4\pi\mu^2}{-s_{qg}}\right)^{\epsilon}
\Biggl\{
\frac{N}{2\epsilon^2}\left[1
-\sum_{m=1}^{\infty} \e^m \left[ \Li_m\left(\OMWW\right)
-\frac{1}{N^2} \Li_m\left(\WW\right)\right]\right]\nonumber \\
&&\qquad\qquad\qquad -\left(\frac{N^2+1}{N}\right)
\frac{w}{4(1-2\e)}
\Biggr\},\\
\lefteqn{r_{S}^{Q \to qg,(1)}(\mp,q^{\pm},g^{\mp})=
r_{S}^{\overline{Q} \to g\bar q,(1)}(\mp,g^{\pm},\bar{q}^{\mp})=}\nonumber \\
&&
-c_{\Gamma}(\e)\left(\frac{4\pi\mu^2}{-s_{qg}}\right)^{\epsilon}
\Biggl\{
\frac{N}{2\epsilon^2}\left[1
-\sum_{m=1}^{\infty} \e^m \left[ \Li_m\left(\OMWW\right)
-\frac{1}{N^2} \Li_m\left(\WW\right)\right]\right]
\Biggr\}. 
\end{eqnarray}
As usual, the polylogarithms ${\rm Li}_n(w)$ are defined by
\begin{eqnarray}
 {\rm Li}_n(w) &=& \int_0^w \frac{dt}{t} {\rm Li}_{n-1}(t) \qquad {\rm ~for~}
 n=3,4,\ldots\nonumber \\
 {\rm Li}_2(w) &=& -\int_0^w \frac{dt}{t} \ln(1-t).
\label{eq:lidef}
\end{eqnarray} 

We have checked that our procedure of taking the collinear limit of the explicit one-loop amplitudes 
written in terms of 2dHPL reproduces these expressions through 
to ${\cal O}(\e^2)$.  Note that,
unlike Ref.~\cite{Bern:split1QCD}, in our notation the gluon always carries a momentum fraction $w$.
Note also that because of our different normalisations, the one-loop splitting function differs by a factor of
$N/2$ compared to Ref.~\cite{Bern:split1QCD}.

\subsection{The two-loop quark-gluon splitting amplitudes}

Explicit calculations for the unrenormalised two-loop coefficients, 
that independently satisfy the pole structure predicted by Catani (see Appendix~\ref{app:catani})
can be found in Ref.~\cite{Garland:2lhelsee->3j}.  Using FORM~\cite{Vermaseren:FORM} and MAPLE to take the
limits of the 2dHPL, 
we find that the unrenormalised splitting functions $r_{S}^{Q \to qg,(2)}(\e)$ 
have a pole structure determined by
Eq.~(\ref{eq:r2}).  The finite remainders are given by,
\begin{eqnarray}
\label{eq:r2qgpp}
\lefteqn{r_{S}^{Q \to qg,(2),fin}(\mp,q^{\pm},g^{\pm})=
r_{S}^{\overline{Q} \to g\bar q,(2),fin}(\mp,g^{\pm},\bar{q}^{\pm})=}\nonumber \\&&
+{N^2}\,\Biggl (-{3\over 2}\,{\Lidx}-{5\over 4}\,{\Lidy}+{3\over 2}\,{\Lidz}+{\Licx}\,{\lx}+{1\over 3}\,{\Licx}+{\Licy}\,{\lx}\nonumber \\&&
+{13\over 24}\,{\Licy}+{1\over 8}\,{\Libx^2}+{1\over 4}\,{\Libx}\,{\lx}\,{\ly}-{1\over 6}\,{\Libx}\,{\lx}+{3\over 4}\,{\Libx}\,{\zeta_2}-{1\over 8}\,{\Libx}\nonumber \\&&
+{1\over 2}\,{\lx^2}\,{\ly^2}-{1\over 4}\,{\lx}\,{\ly^3}-{3\over 4}\,{\lx}\,{\ly}\,{\zeta_2}-{1\over 8}\,{\lx}\,{\ly}-{\lx}\,{\zeta_3}\nonumber \\&&
+{1\over 16}\,{\ly^4}+{3\over 4}\,{\ly^2}\,{\zeta_2}+{1\over 4}\,{\ly}\,{\zeta_3}+{55\over 48}\,{\ly}\,{\zeta_2}-{193\over 108}\,{\ly}+{29\over 32}\,{\zeta_4}\nonumber \\&&
+{527\over 144}\,{\zeta_3}+{335\over 288}\,{\zeta_2}-{571\over 324}\Biggr )\nonumber \\&&
+\,\Biggl (-{2}\,{\Lidx}-{7\over 4}\,{\Lidy}+{\Lidz}+{\Licx}\,{\lx}+{1\over 4}\,{\Licx}\,{\ly}+{3\over 2}\,{\Licx}\nonumber \\&&
+{1\over 2}\,{\Licy}\,{\lx}+{1\over 2}\,{\Licy}\,{\ly}+{9\over 4}\,{\Licy}-{1\over 8}\,{\Libx^2}-{1\over 4}\,{\Libx}\,{\lx}\,{\ly}\nonumber \\&&
-{3\over 4}\,{\Libx}\,{\lx}+{3\over 4}\,{\Libx}\,{\zeta_2}-{3\over 4}\,{\Liby}\,{\ly}+{1\over 4}\,{\lx^2}\,{\ly^2}-{1\over 6}\,{\lx}\,{\ly^3}\nonumber \\&&
-{1\over 4}\,{\lx}\,{\ly}\,{\zeta_2}-{1\over 2}\,{\lx}\,{\zeta_3}+{1\over 24}\,{\ly^4}+{1\over 2}\,{\ly^2}\,{\zeta_2}+{3\over 4}\,{\ly}\,{\zeta_3}\nonumber \\&&
+{11\over 48}\,{\ly}\,{\zeta_2}-{323\over 216}\,{\ly}+{7\over 4}\,{\zeta_4}-{9\over 4}\,{\zeta_3}\nonumber \\&&
+{w}\,\Biggl (-{3\over 4}\,{\Licx}-{3\over 4}\,{\Licy}+{1\over 4}\,{\Libx}\,{\lx}+{1\over 4}\,{\Libx}+{1\over 4}\,{\Liby}\,{\ly}\nonumber \\&&
+{1\over 8}\,{\lx}\,{\ly}+{1\over 4}\,{\lx}\,{\zeta_2}+{3\over 8}\,{\lx}+{1\over 4}\,{\ly}\,{\zeta_2}-{3\over 8}\,{\ly}-{1\over 8}\,{\zeta_2}+{3\over 32}\Biggr )\nonumber \\&&
+{w^2}\,\Biggl (-{3\over 8}\,{\Licx}-{3\over 8}\,{\Licy}+{1\over 8}\,{\Libx}\,{\lx}+{1\over 8}\,{\Liby}\,{\ly}+{1\over 8}\,{\lx}\,{\zeta_2}\nonumber \\&&
+{1\over 8}\,{\ly}\,{\zeta_2}-{1\over 32}\Biggr )\Biggr )\nonumber \\&&
+{1\over N^2}\Biggl (-{1\over 2}\,{\Lidx}-{1\over 2}\,{\Lidy}-{1\over 2}\,{\Lidz}+{1\over 4}\,{\Licx}\,{\ly}-{1\over 2}\,{\Licy}\,{\lx}\nonumber \\&&
+{1\over 2}\,{\Licy}\,{\ly}+{3\over 8}\,{\Licy}-{1\over 4}\,{\Libx^2}-{1\over 2}\,{\Libx}\,{\lx}\,{\ly}+{1\over 8}\,{\Libx}\nonumber \\&&
-{1\over 4}\,{\lx^2}\,{\ly^2}+{1\over 12}\,{\lx}\,{\ly^3}+{1\over 2}\,{\lx}\,{\ly}\,{\zeta_2}+{1\over 8}\,{\lx}\,{\ly}\nonumber \\&&
+{1\over 2}\,{\lx}\,{\zeta_3}-{1\over 48}\,{\ly^4}-{1\over 4}\,{\ly^2}\,{\zeta_2}+{1\over 2}\,{\ly}\,{\zeta_3}-{3\over 4}\,{\ly}\,{\zeta_2}+{3\over 8}\,{\ly}\nonumber \\&&
+{1\over 2}\,{\zeta_4}-{3\over 8}\,{\zeta_3}\nonumber \\&&
+{w}\,\Biggl (-{3\over 4}\,{\Licx}-{3\over 4}\,{\Licy}+{1\over 4}\,{\Libx}\,{\lx}+{1\over 4}\,{\Libx}+{1\over 4}\,{\Liby}\,{\ly}\nonumber \\&&
+{1\over 8}\,{\lx}\,{\ly}+{1\over 4}\,{\lx}\,{\zeta_2}+{3\over 8}\,{\lx}+{1\over 4}\,{\ly}\,{\zeta_2}-{3\over 8}\,{\ly}-{1\over 8}\,{\zeta_2}+{3\over 32}\Biggr )\nonumber \\&&
+{w^2}\,\Biggl (-{3\over 8}\,{\Licx}-{3\over 8}\,{\Licy}+{1\over 8}\,{\Libx}\,{\lx}+{1\over 8}\,{\Liby}\,{\ly}+{1\over 8}\,{\lx}\,{\zeta_2}\nonumber \\&&
+{1\over 8}\,{\ly}\,{\zeta_2}-{1\over 32}\Biggr )\Biggr )\nonumber \\&&
+{N}\,{\NF}\,\Biggl (-{1\over 3}\,{\Licx}-{1\over 6}\,{\Licy}+{1\over 6}\,{\Libx}\,{\lx}-{5\over 24}\,{\ly}\,{\zeta_2}+{19\over 108}\,{\ly}\nonumber \\&&
-{43\over 72}\,{\zeta_3}-{25\over 144}\,{\zeta_2}+{65\over 648}\Biggr )
+{\NF\over N}\,{}\,\Biggl (-{1\over 24}\,{\ly}\,{\zeta_2}+{7\over 27}\,{\ly}\Biggr )\nonumber \\&&
+w\left(\frac{\zeta_2}{4}N^2+\frac{1}{24}\frac{\NF}{N}
+\frac{5}{24}N(N-\NF)+\left(\frac{\zeta_2}{4}-\frac{1}{24}\right)
\right)+\frac{w(1-w)}{32}(N^2+1),
\end{eqnarray}
\begin{eqnarray}
\label{eq:r2qgpm}
\lefteqn{r_{S}^{Q \to qg,(2),fin}(\mp,q^{\pm},g^{\mp})=
r_{S}^{\overline{Q} \to g\bar q,(2),fin}(\mp,g^{\pm},\bar{q}^{\mp})=
}\nonumber \\&&
+{N^2}\,\Biggl (-{3\over 2}\,{\Lidx}-{5\over 4}\,{\Lidy}+{3\over 2}\,{\Lidz}+{\Licx}\,{\lx}+{1\over 3}\,{\Licx}+{\Licy}\,{\lx}\nonumber \\&&
+{13\over 24}\,{\Licy}+{1\over 8}\,{\Libx^2}+{1\over 4}\,{\Libx}\,{\lx}\,{\ly}-{1\over 6}\,{\Libx}\,{\lx}+{3\over 4}\,{\Libx}\,{\zeta_2}\nonumber \\&&
-{1\over 8}\,{\Libx}+{1\over 2}\,{\lx^2}\,{\ly^2}-{1\over 4}\,{\lx}\,{\ly^3}-{3\over 4}\,{\lx}\,{\ly}\,{\zeta_2}-{1\over 8}\,{\lx}\,{\ly}\nonumber \\&&
-{\lx}\,{\zeta_3}+{1\over 12}\,{\lx}+{1\over 16}\,{\ly^4}+{3\over 4}\,{\ly^2}\,{\zeta_2}+{1\over 4}\,{\ly}\,{\zeta_3}+{55\over 48}\,{\ly}\,{\zeta_2}\nonumber \\&&
-{193\over 108}\,{\ly}+{29\over 32}\,{\zeta_4}+{527\over 144}\,{\zeta_3}+{335\over 288}\,{\zeta_2}-{571\over 324}\Biggr )\nonumber \\&&
+\,\Biggl (-{2}\,{\Lidx}-{7\over 4}\,{\Lidy}+{\Lidz}+{\Licx}\,{\lx}+{1\over 4}\,{\Licx}\,{\ly}+{3\over 2}\,{\Licx}\nonumber \\&&
+{1\over 2}\,{\Licy}\,{\lx} +{1\over 2}\,{\Licy}\,{\ly}+{3\over 4}\,{\Licy}-{1\over 8}\,{\Libx^2}-{1\over 4}\,{\Libx}\,{\lx}\,{\ly}\nonumber \\&&
-{3\over 4}\,{\Libx}\,{\lx}+{3\over 4}\,{\Libx}\,{\zeta_2}+{1\over 4}\,{\lx^2}\,{\ly^2}-{1\over 6}\,{\lx}\,{\ly^3}-{1\over 4}\,{\lx}\,{\ly}\,{\zeta_2}\nonumber \\&&
-{1\over 2}\,{\lx}\,{\zeta_3}-{3\over 8}\,{\lx}+{1\over 24}\,{\ly^4}+{1\over 2}\,{\ly^2}\,{\zeta_2}+{3\over 4}\,{\ly}\,{\zeta_3}-{25\over 48}\,{\ly}\,{\zeta_2}\nonumber \\&&
-{323\over 216}\,{\ly}+{7\over 4}\,{\zeta_4}-{3\over 4}\,{\zeta_3}\Biggr )\nonumber \\&&
+{1\over N^2}\,\Biggl (-{1\over 2}\,{\Lidx}-{1\over 2}\,{\Lidy}-{1\over 2}\,{\Lidz}+{1\over 4}\,{\Licx}\,{\ly}-{1\over 2}\,{\Licy}\,{\lx}\nonumber \\&&
+{1\over 2}\,{\Licy}\,{\ly}+{3\over 8}\,{\Licy}-{1\over 4}\,{\Libx^2}-{1\over 2}\,{\Libx}\,{\lx}\,{\ly}+{1\over 8}\,{\Libx}\nonumber \\&&
-{1\over 4}\,{\lx^2}\,{\ly^2}+{1\over 12}\,{\lx}\,{\ly^3}+{1\over 2}\,{\lx}\,{\ly}\,{\zeta_2}+{1\over 8}\,{\lx}\,{\ly}+{1\over 2}\,{\lx}\,{\zeta_3}\nonumber \\&&
-{3\over 8}\,{\lx}-{1\over 48}\,{\ly^4}-{1\over 4}\,{\ly^2}\,{\zeta_2}+{1\over 2}\,{\ly}\,{\zeta_3}-{3\over 4}\,{\ly}\,{\zeta_2}+{3\over 8}\,{\ly}\nonumber \\&&
+{1\over 2}\,{\zeta_4}-{3\over 8}\,{\zeta_3}\Biggr )\nonumber \\&&
+{N}\,{\NF}\,\Biggl (-{1\over 3}\,{\Licx}-{1\over 6}\,{\Licy}+{1\over 6}\,{\Libx}\,{\lx}-{1\over 12}\,{\lx}-{5\over 24}\,{\ly}\,{\zeta_2}\nonumber \\&&
+{19\over 108}\,{\ly}-{43\over 72}\,{\zeta_3}-{25\over 144}\,{\zeta_2}+{65\over 648}\Biggr )\nonumber \\&&
+{\NF\over N}\,\Biggl (-{1\over 24}\,{\ly}\,{\zeta_2}+{7\over 27}\,{\ly}\Biggr )\nonumber \\&&
-N^2\left(\frac{w(4-3w)}{4(1-w)^2}\left(\AAAXXX\right)
+\frac{w}{8(1-w)}\Biggl(\AAAYYY\Biggr)\right)\nonumber \\&&
+\frac{1}{N^2}\left(\frac{w(3w-2)}{4(1-w)^2}\left(\AAAXXX\right)
+\frac{w}{8(1-w)}\Biggl(\AAAYYY\Biggr)\right)\nonumber \\&&
-N(N-\NF)\frac{1}{12}\left(\frac{\lx}{(1-w)}\right)
+\left(\frac{N^2+1}{N^2}\right)\frac{3}{8}\left(\frac{\lx}{(1-w)}\right).
\end{eqnarray}

\section{The gluon-gluon splitting amplitudes}
\label{sec:gluongluon}

The gluon-gluon splitting amplitudes can be extracted from
the Higgs decay into gluons in the limit where the top quark is very heavy and an 
effective Higgs-gluon-gluon vertex is induced.
Labelling the gluon momenta alphabetically we write the unrenormalised amplitude in terms of
the single colour ordered amplitude,
\begin{equation}
|{\cal M}_{H \to A B }\rangle = {\bom \delta_{AB}} \,{\cal A}_H(A,B),
\end{equation}
where,
\begin{eqnarray}
\label{eq:helampHgg}
{\cal A}_H(A^+,B^+)&=& A_H ~[A B]^2,\\
{\cal A}_H(A^+,B^-) &=& 0.
\end{eqnarray}
The amplitudes where the first gluon has negative helicity can be obtained by
parity.

Similarly, we write the Higgs to three gluon amplitude as,
\begin{equation}
|{\cal M}_{H \to a b c} \rangle = {\bom f^{abc}} \, {\cal A}_H(a,b,c).
\end{equation}
Again, there is only one colour ordering 
and the two independent helicity amplitudes for the three gluon decay are~\cite{Schmidt:1997wr,inprep},
\begin{eqnarray}
\label{eq:helamphggg1}
{\cal A}_H( a^+ ,b^+ ,c^+)  &=& -\alpha_H(y,z)
~\frac{1}{\sqrt{2}}\frac{s_{abc}^2}{\langle ab\rangle\langle bc\rangle\langle ac\rangle}
,\\
\label{eq:helamphggg2}
{\cal A}_H(a^+,b^-,c^+) &=& \beta_H(y,z)
~\frac{1}{\sqrt{2}}\frac{[ac]^4}{[ab][bc][ca]}, 
\end{eqnarray}
where $y = s_{ab}/s_{abc}$ and $z = s_{ac}/s_{abc}$.

The other helicity amplitudes are obtained by the usual parity and charge
conjugation relations.  For the present purposes, the only useful one is,
\begin{eqnarray}
\label{eq:helamphggg3}
{\cal A}_{H}(a^-, b^-, c^+) &=& 
\beta_H(z,y)~\frac{1}{\sqrt{2}}\frac{\langle ab \rangle^4}{\langle ab
\rangle\langle bc \rangle\langle ca \rangle}. 
\end{eqnarray}

The limit where gluons $a$ and $b$ are collinear corresponds
to $y = s_{ab}/s_{abc} \to 0$.  We assign a momentum fraction $(1-w)$ to particle $a$.  In this limit,
we see that,
\begin{eqnarray}
{\cal A}_{H}(a^+,b^+,c^+) &\stackrel{a||b}{\longrightarrow}&
 \frac{1}{\sqrt{2}} 
 \frac{1}{\sqrt{w(1-w)}\langle ab\rangle}
 \biggl( \alpha_H(y,z)\Big|_{y \to 0}~\biggr)[A B]^2,\nonumber \\
{\cal A}_{H}(a^+ ,b^- ,c^+) &\stackrel{a||b}{\longrightarrow}& 
-\frac{1}{\sqrt{2}}\frac{(1-w)^2}{\sqrt{w(1-w)}[ab]}
 \biggl( \beta_H(y,z)\Big|_{y \to 0}~\biggr)[A B]^2,\nonumber \\
{\cal A}_{H}(a^-, b^- ,c^+) &\stackrel{a||b}{\longrightarrow}& 
-\frac{1}{\sqrt{2}}\frac{\langle ab\rangle}{\sqrt{w(1-w)}[ab]^2}
\biggl(\frac{s_{ab}^2}{s_{abc}^2} \beta_H(z,y)\Big|_{y \to 0}~\biggr)[A B]^2.
\end{eqnarray}
Note that the terms in round brackets are evaluated in the collinear limit, $y \to 0$.

The two-particle helicity amplitude coefficient $A_H$ 
has the perturbative expansion,
\begin{equation}
A_H =    C_{H \to GG}\, \left[
A_H^{(0)}  
+ \left(\frac{\alpha_s}{2\pi}\right) A_H^{(1)}  
+ \left(\frac{\alpha_s}{2\pi}\right)^2 A_H^{(2)} 
+ {\cal O}(\alpha_s^3) \right] .
\end{equation}
Similarly, the three-particle helicity amplitude coefficients $\alpha_H$,
and $\beta_H$ are given by, 
\begin{equation}
\Omega_H(y,z) =  C_{H\to ggg}\, \left[
\Omega_H^{(0)}(y,z)  
+ \left(\frac{\alpha_s}{2\pi}\right) \Omega_H^{(1)}(y,z)  
+ \left(\frac{\alpha_s}{2\pi}\right)^2 \Omega_H^{(2)}(y,z) 
+ {\cal O}(\alpha_s^3) \right] \;,\nonumber \\
\end{equation}
for $\Omega_H = \alpha_H,\beta_H$.

\subsection{The tree-level gluon-gluon splitting amplitudes}

At leading order 
\begin{equation}
\alpha_H^{(0)}(y,z) = \beta_H^{(0)}(y,z) =  1\qquad
\mbox{and}\qquad A_H^{(0)}=1,
\end{equation}
and we can immediately read off the tree-level gluon-gluon 
splitting functions,
\begin{eqnarray}
\Split_{-}^{G \to gg,(0)}(a^{+},b^{+})&=&
\frac{1}{\sqrt{2}} \frac{1}{\sqrt{w(1-w)} ~\langle ab \rangle},\nonumber \\
\Split_{-}^{G \to gg,(0)}(a^{+},b^{-})&=&
-\frac{1}{\sqrt{2}} \frac{(1-w)^2}{\sqrt{w(1-w)} ~[ ab ]},\nonumber \\
\Split_{-}^{G \to gg,(0)}(a^{-},b^{-})&=& 0,
\end{eqnarray}
and by symmetry,
\begin{eqnarray}
\Split_{-}^{G \to gg,(0)}(a^{-},b^{+})&=&
-\frac{1}{\sqrt{2}} \frac{w^2}{\sqrt{w(1-w)} ~[ ab ]}.
\end{eqnarray}
The amplitudes where a gluon with negative helicity splits are obtained by parity,
\begin{eqnarray}
\Split_{+}^{G \to gg,(0)}(a^{\lambda_a},b^{\lambda_b})&=&
\left(\Split_{-}^{G \to gg,(0)}(a^{-\lambda_a},b^{-\lambda_b})\right)^*,
\end{eqnarray}
and recalling that $[ ab ]^* = -\langle ab\rangle$.

\subsection{The one-loop gluon-gluon splitting amplitudes}

As in the quark-gluon case, the one loop splitting functions are presented in 
Ref.~\cite{Bern:split1QCD}, and we
reproduce them here in the 't Hooft-Veltman regularisation scheme
($\delta_R = 1$) for the sake of completeness.
For the amplitude that vanishes at tree-level we have,
\begin{eqnarray}
\Split_{-}^{G \to gg,(1)}(a^{-},b^{-})&=&c_\Gamma
\left(\frac{4\pi\mu^2}{-s_{ab}}\right)^{\epsilon}
\frac{w(1-w)}{(1-2\e)(2-2\e)(3-2\e)}\left( N-N\e-\NF\right)\nonumber
\\
&\times& -\frac{1}{\sqrt{2}}\frac{\langle ab\rangle}{\sqrt{w(1-w)}[ab]^2},
\end{eqnarray}
while the other splitting functions are,
\begin{eqnarray}
\label{eq:r1ggpp}
\lefteqn{r_{S}^{G \to gg,(1)}(\mp,a^{\pm},b^{\pm})=}\nonumber \\
&&
c_{\Gamma}(\e)\left(\frac{4\pi\mu^2}{-s_{ab}}\right)^{\epsilon}
\Biggl\{
\frac{N}{2\epsilon^2}\left[
-\Gamma(1-\epsilon)\Gamma(1+\epsilon)\left(\frac{w}{1-w}\right)^{\epsilon}
+\sum_{m=1}^{\infty} 2\epsilon^{2m-1} \Li_{2m-1}\left(\OMWW\right)
\right]\nonumber \\
&&\qquad\qquad\qquad+ \frac{w(1-w)}{(1-2\e)(2-2\e)(3-2\e)}\left(N-N\e-\NF\right)
\Biggr\},\\
\label{eq:r1ggpm}
\lefteqn{r_{S}^{G \to gg,(1)}(\mp,a^{\pm},c^{\mp})=r_{S}^{G \to gg,(1)}(\mp,a^{\mp},c^{\pm})=}\nonumber \\
&&
c_{\Gamma}(\e)\left(\frac{4\pi\mu^2}{-s_{ab}}\right)^{\epsilon}
\frac{N}{2\epsilon^2}\left[
-\Gamma(1-\epsilon)\Gamma(1+\epsilon)\left(\frac{w}{1-w}\right)^{\epsilon}
+\sum_{m=1}^{\infty} 2\epsilon^{2m-1} \Li_{2m-1}\left(\OMWW\right)
\right].\nonumber \\
\end{eqnarray}
Note that Eqs.~(\ref{eq:r1ggpp}) and (\ref{eq:r1ggpm}) are symmetric under $w \leftrightarrow (1-w)$.
Note also that because of our different normalisations, the one-loop splitting function differs by a factor of
$N/2$ compared to Ref.~\cite{Bern:split1QCD}. 
We have checked that we reproduce these expressions through 
to ${\cal O}(\e^2)$.
 
\subsection{The two-loop gluon-gluon splitting amplitudes}

Using explicit calculations for the unrenormalised two-loop coefficients for the $H \to gg$ and
$H \to ggg$ decays, 
we find that the unrenormalised splitting function $r_{S}^{G \to gg,(2)}(\e)$ for the   
helicity combinations that are non-vanishing at tree level 
have a pole structure determined by
Eq.~(\ref{eq:r2}) with finite remainders given by,
\begin{eqnarray}
\label{eq:r2ggpp}
\lefteqn{r_{S}^{G \to gg,(2),fin}(-,g^{+},g^{+})=}\nonumber \\&&
+{N^2}\,\Biggl (-{11\over 32}\,{\zeta_4}+{5\over 24}\,{\zeta_2}-{523\over 432}\Biggr )\nonumber \\&&
+{\beta_0}\,{N}\,\Biggl (-{\Licx}-{\Licy}+{1\over 2}\,{\Libx}\,{\lx}+{1\over 2}\,{\Liby}\,{\ly}+{79\over 24}\,{\zeta_3}+{25\over 48}\,{\zeta_2}-{65\over 216}\Biggr )
\nonumber \\&&
+w(1\!-\!w)
\left(\frac{47}{108}N^2
-\frac{23}{216}NN_F
-\frac{11}{54}N_F^2
-\frac{1}{8}\frac{N_F}{N}
+\frac{N(N\!-\!N_F)}{12}\left(2\zeta_2+\frac{\ln(w)}{(1\!-\!w)}+\frac{\ln(1\!-\!w)}{w}\right)
\right)\nonumber \\&&
-w^2(1-w)^2(N-N_F)^2 \frac{1}{72},\\&&
\label{eq:r2ggpm}
\lefteqn{r_{S}^{G \to gg,(2),fin}(-,g^{+},g^{-})=}\nonumber \\&&
+N^2 \,\Biggl (-{3\over 8}\,{\lx}-{3\over 8}\,{\ly}-{11\over 32}\,{\zeta_4}+{5\over 24}\,{\zeta_2}-{523\over 432}\Biggr )\nonumber \\&&
+{\beta_0}\,{N}\,\Biggl (-{\Licx}+{1\over 2}\,{\Libx}\,{\lx}+{1\over 4}\,{\lx}+{1\over 2}\,{\ly}\,{\zeta_2}+{1\over 4}\,{\ly}\nonumber \\&&
+{55\over 24}\,{\zeta_3}+{25\over 48}\,{\zeta_2}-{65\over 216}\Biggr )
\nonumber \\&&
-N\beta_0 \frac{w}{(1-w)}\left(\AAAXXX\right)\nonumber \\&&
-N(N-\NF)\Biggl (
\frac{w(1+w)}{6(1-w)^3}\left(\AAAXXX\right)\nonumber \\&&
+\frac{w}{6(1-w)^2}\Biggl(\AAAYYY\Biggr)
+\frac{1}{12}\left(\frac{\lx}{(1-w)}\right)\Biggr).
\end{eqnarray}
Finally, for the amplitude that vanishes at tree-level we find,
\begin{eqnarray}
\label{eq:r2ggmm}
\Split_{-}^{G \to gg,(2)}(a^{-},b^{-})&=&
c_{\Gamma}(\e)\left(\frac{4\pi\mu^2}{-s_{ab}}\right)^{\epsilon}
\left( \frac{N}{2\e^2} (w(1-w))^{-\e} - \frac{2\beta_0}{\e}\right)\,\Split_{-}^{G \to gg,(1)}(a^{-},b^{-})\nonumber \\
-\frac{1}{\sqrt{2}}\frac{\langle ab\rangle}{\sqrt{w(1-w)}[ab]^2}&\times&
\Biggr[N\,(N-\NF)\,\left(\frac{1}{6}\ly
+w\, \frac{1}{6}\, (\lx-\ly)\right)\nonumber \\
&&-w\,(1-w) \Biggl(\left(-\frac{101}{54}+\frac{\zeta_2}{12}-\frac{1}{6}\lx\ly\right)N^2 \nonumber \\&&
+\left(\frac{451}{216}-\frac{\zeta_2}{12}+\frac{1}{6}\lx\ly\right)N \NF 
+ \frac{1}{8} \frac{\NF}{N} -\frac{5}{54} \NF^2\Biggr)\Biggr]+{\cal O}(\e). \nonumber \\
\end{eqnarray}
We have checked that the expressions presented here agree with those given in
Ref.~\cite{Bern:2lsplit}.

\section{The quark-antiquark splitting amplitudes}
\label{sec:quarkantiquark}

The quark-antiquark
splitting amplitudes can be extracted from
the Higgs decay into quarks and gluons in the large $m_t$ limit, which factorises onto the $H \to GG$ amplitudes in
the collinear limit. 
The single colour ordered amplitude is obtained from the 
full amplitude by,
\begin{equation}
|{\cal M}_{H \to q  \bar q g}\rangle = {\bom T^c_{ab}} \, {\cal A}_H(q,\bar
q,g).
\end{equation}

There 
is one independent helicity amplitude for the $H \to qg\bar q$ decay 
which is given by~\cite{Schmidt:1997wr,inprep},
\begin{eqnarray}
\label{eq:helamphqqg}
{\cal A}_H(q^+,\bar q^-, g^+) &=& \gamma_H(y,z)~\frac{1}{\sqrt{2}}
\frac{[qg]^2}{[q\bar q]},
\end{eqnarray}
where $y = s_{q\bar q}/s_{q\bar qg}$ and $z = s_{qg}/s_{q\bar qg}$.
All other helicity amplitudes are either zero due to helicity conservation 
or can be related by charge conjugation and parity.

In the collinear quark and antiquark limit, $y = s_{q\bar q}/s_{q\bar qg} \to 0$.
When the quark carries momentum fraction $(1-w)$) we see that,
\begin{eqnarray}
{\cal A}_H(q^+,\bar q^-, g^+) &\stackrel{q||\bar q}{\longrightarrow}&
\frac{1}{\sqrt{2}} 
 \frac{w}{[q\bar q]}\biggl( \gamma_H(y,z)\Big|_{y \to 0}~\biggr)[A B]^2.
\end{eqnarray}
Note that the terms in round brackets are evaluated in the collinear limit, $y \to 0$.

The three-particle helicity amplitude coefficient $\gamma_H$,
has a perturbative expansion 
given by, 
\begin{equation}
\gamma_H(y,z) =  C_{H \to q\bar q g} \left[
\gamma_H^{(0)}(y,z)  
+ \left(\frac{\alpha_s}{2\pi}\right) \gamma_H^{(1)}(y,z)  
+ \left(\frac{\alpha_s}{2\pi}\right)^2 \gamma_H^{(2)}(y,z) 
+ {\cal O}(\alpha_s^3) \right] .
\end{equation}

\subsection{The tree-level quark-antiquark splitting amplitudes}

At leading order 
\begin{equation}
\gamma_H^{(0)} = 1.
\end{equation}
Together with the tree amplitude for $H \to GG$ in Eq.~(\ref{eq:helampHgg}), 
we can immediately read off the tree-level quark-antiquark 
splitting function,
\begin{eqnarray}
\Split_{-}^{G \to q\bar q,(0)}(q^{+},\bar{q}^{-})&=&
\frac{1}{\sqrt{2}} \frac{w}{[ q\bar q ]}. 
\end{eqnarray}

\subsection{The one-loop quark-antiquark splitting amplitudes}

As in the quark-gluon and gluon-gluon
case, the one loop splitting functions are presented in 
Ref.~\cite{Bern:split1QCD}. In the 't Hooft-Veltman regularisation scheme
($\delta_R = 1$),
\begin{eqnarray}
\lefteqn{r_{S}^{G \to q\bar q,(1)}(-,q^{+},\bar{q}^{-};\e)=}\nonumber \\
&&
c_{\Gamma}(\e)\left(\frac{4\pi\mu^2}{-s_{q\bar q}}\right)^{\epsilon}
\Biggl\{
\frac{N}{2\epsilon^2}\left[
1-\Gamma(1-\epsilon)\Gamma(1+\epsilon)\left(\frac{w}{1-w}\right)^{\epsilon}
+\sum_{m=1}^{\infty} 2\epsilon^{2m-1} \Li_{2m-1}\left(\OMWW\right)
\right]\nonumber \\
&&\qquad\qquad\qquad+
N\left(\frac{13}{12\e(1-2\e)}+\frac{1}{6(1-2\e)(3-2\e)}\right)
+\frac{1}{N}\left(\frac{1}{2\e^2}+\frac{3}{4\e(1-2\e)}+\frac{1}{2(1-2\e)}\right)\nonumber
\\
&&\qquad\qquad\qquad+\NF\left(-\frac{1}{3\e(1-2\e)}+\frac{1}{3(1-2\e)(3-2\e)}\right)
\Biggr\}.
\end{eqnarray}
Note that because of our different normalisations, the one-loop splitting function differs by a factor of
$N/2$ compared to Ref.~\cite{Bern:split1QCD}. 
We have checked that we agree with this expression through 
to ${\cal O}(\e^2)$.

\subsection{The two-loop quark-antiquark splitting amplitudes}

Using explicit calculations for the unrenormalised two-loop coefficients for the $H \to gg$ and
$H \to q\bar q g$ processes, 
we find that the unrenormalised splitting function $r_{S}^{G \to q\bar q,(2)}(\e)$ has 
a pole structure determined by
Eq.~(\ref{eq:r2}) with a finite remainder given by,
\begin{eqnarray}
\label{eq:r2qqpp}
\lefteqn{r_{S}^{G \to q\bar q,(2),fin}(-,q^{+},\bar{q}^{+})=}\nonumber \\&&
+{N^2}\,\Biggl ({5\over 4}\,{\Lidx}+{5\over 4}\,{\Lidy}-{1\over 2}\,{\Licx}\,{\lx}-{3\over 4}\,{\Licx}\,{\ly}-{3\over 4}\,{\Licx}\nonumber \\&&
-{3\over 4}\,{\Licy}\,{\lx}-{1\over 2}\,{\Licy}\,{\ly}-{1\over 4}\,{\Libx^2}-{1\over 4}\,{\Libx}\,{\lx}\,{\ly}\nonumber \\&&
+{3\over 8}\,{\Libx}\,{\lx}+{1\over 4}\,{\Libx}\,{\zeta_2}+{1\over 8}\,{\Libx}-{3\over 8}\,{\lx^2}\,{\ly^2}+{3\over 4}\,{\lx}\,{\ly}\,{\zeta_2}\nonumber \\&&
+{1\over 8}\,{\lx}\,{\ly}+{5\over 4}\,{\lx}\,{\zeta_3}-{29\over 48}\,{\lx}\,{\zeta_2}+{121\over 864}\,{\lx}+{5\over 4}\,{\ly}\,{\zeta_3}-{11\over 48}\,{\ly}\,{\zeta_2}\nonumber \\&&
+{121\over 864}\,{\ly}+{23\over 16}\,{\zeta_4}+{5\over 24}\,{\zeta_3}-{31\over 96}\,{\zeta_2}-{14233\over 3456}\Biggr )\nonumber \\&&
+\,\Biggl ({5\over 4}\,{\Lidx}+{5\over 4}\,{\Lidy}-{1\over 2}\,{\Licx}\,{\lx}-{3\over 4}\,{\Licx}\,{\ly}+{3\over 4}\,{\Licx}\nonumber \\&&
-{3\over 4}\,{\Licy}\,{\lx}-{1\over 2}\,{\Licy}\,{\ly}-{1\over 4}\,{\Libx^2}-{1\over 4}\,{\Libx}\,{\lx}\,{\ly}\nonumber \\&&
-{3\over 8}\,{\Libx}\,{\lx}+{1\over 4}\,{\Libx}\,{\zeta_2}+{1\over 8}\,{\Libx}-{3\over 8}\,{\lx^2}\,{\ly^2}+{3\over 4}\,{\lx}\,{\ly}\,{\zeta_2}\nonumber \\&&
+{1\over 8}\,{\lx}\,{\ly}-{1\over 4}\,{\lx}\,{\zeta_3}+{5\over 16}\,{\lx}\,{\zeta_2}-{41\over 108}\,{\lx}-{1\over 4}\,{\ly}\,{\zeta_3}-{1\over 16}\,{\ly}\,{\zeta_2}\nonumber \\&&
-{41\over 108}\,{\ly}+{13\over 32}\,{\zeta_4}-{749\over 144}\,{\zeta_3}+{47\over 48}\,{\zeta_2}-{11389\over 2592}\Biggr)\nonumber \\&&
+{1\over N^2}\,\Biggl (-{3\over 2}\,{\lx}\,{\zeta_3}+{3\over 4}\,{\lx}\,{\zeta_2}-{3\over 32}\,{\lx}-{3\over 2}\,{\ly}\,{\zeta_3}+{3\over 4}\,{\ly}\,{\zeta_2}-{3\over 32}\,{\ly}\nonumber \\&&
-{11\over 8}\,{\zeta_4}-{15\over 8}\,{\zeta_3}+{29\over 16}\,{\zeta_2}-{1\over 128}\Biggr )\nonumber \\&&
+{N}\,{\NF}\,\Biggl ({1\over 24}\,{\lx}\,{\zeta_2}+{5\over 8}\,{\lx}+{1\over 24}\,{\ly}\,{\zeta_2}+{5\over 8}\,{\ly}-{5\over 6}\,{\zeta_3}+{31\over 144}\,{\zeta_2}+{1717\over 864}\Biggr )\nonumber \\&&
+{\NF\over N}\, \Biggl (-{1\over 8}\,{\lx}\,{\zeta_2}+{79\over 216}\,{\lx}-{1\over 8}\,{\ly}\,{\zeta_2}+{79\over 216}\,{\ly}-{53\over 72}\,{\zeta_3}-{7\over 24}\,{\zeta_2}+{5221\over 2592}\Biggr )\nonumber \\&&
+{\NF^2}\,\Biggl (-{5\over 27}\,{\lx}-{5\over 27}\,{\ly}-{1\over 36}\,{\zeta_2}-{1\over 2}\Biggr )\nonumber \\&&
+N^2\left(\frac{w(3w-2)}{4(1-w)^2}\left(\AAAXXX\right)
+\frac{w}{8(1-w)}\Biggl(\AAAYYY\Biggr)\right)\nonumber \\&&
+\left(\frac{w(4-3w)}{4(1-w)^2}\left(\AAAXXX\right)
+\frac{w}{8(1-w)}\Biggl(\AAAYYY\Biggr)\right).
\end{eqnarray}

\section{Limiting Properties}
\label{sec:limits}

It is useful consider various limiting cases of the splitting functions in
order to get a more physical picture of what is going on. 

\DOUBLETABLE[h!]{\label{tab:wlims0}\begin{tabular}{|c|c|c|c|}
	\hline 
	Process & $\Split^{(0)}$ & $r^{(1)}$ & $r^{(2)}$ \\
	\hline 
	$Q^+ \to q^+g^+$ & $\frac{1}{\sqrt{w}}$ & $a_1$ & $a_2$ \\
	\hline 
	$Q^+ \to q^+g^-$ & $\frac{1}{\sqrt{w}}$ & $a_1$ & $a_2$  \\
	\hline 
	$G^+ \to g^+g^+$ & $\frac{1}{\sqrt{w}}$ & $a_1$ & $a_2$  \\
	\hline 
	$G^+ \to g^+g^-$ & $\frac{1}{\sqrt{w}}$ & $a_1$ & $a_2$  \\
	\hline 
	$G^+ \to q^+\bar{q}^+$ & $w$ & $b_1$ & $b_2$ \\
	\hline
	\multicolumn{4}{c}{} \\ 
	\hline 
	Process & $\Split^{(0)}$ & $\Split^{(1)}$ & $\Split^{(2)}$ \\
	\hline
	$G^+ \to g^- g^-$ & 0 & $\sqrt{w}$ & $\sqrt{w}$ \\
	\hline
	\end{tabular}
	}
	{\label{tab:wlims1}\begin{tabular}{|c|c|c|c|}
	\hline 
	Process & $\Split^{(0)}$ & $r^{(1),fin}$ & $r^{(2),fin}$ \\
	\hline 
	$Q^+ \to q^+g^+$ & 1 & $c_1$ & $c_2$ \\
	\hline 
	$Q^+ \to q^+g^-$ & $1-w$ & $a_1$  & $d_2$  \\
	\hline 
	$G^+ \to g^+g^+$ & $\frac{1}{\sqrt{1-w}}$ & $a_1$ & $a_2$ \\
	\hline 
	$G^+ \to g^+g^-$ & $\frac{(1-w)^2}{\sqrt{1-w}}$ & $a_1$  & $\frac{1}{1-w}$ \\
	\hline 
	$G^+ \to q^+\bar{q}^+$ & 1 & $b_1$  & $b_2$ \\ \hline
	\multicolumn{4}{c}{} \\ 
	\hline
	Process & $\Split^{(0)}$ & $\Split^{(1)}$ & $\Split^{(2)}$ \\
	\hline
	$G^+ \to g^- g^-$ & 0 & $\sqrt{1-w}$ & $\sqrt{1-w}$ \\
	\hline
	\end{tabular}
	}{Leading behaviour of splitting functions as $w \to 0$}
	 {Leading behaviour of splitting functions as $w \to 1$}

In Tables~1 and 2 we show the leading behaviour of the splitting functions as a polynomial 
in $w$ for $w \to 0$ and $(1-w)$ for $w \to 1$.
$a_i,b_i,c_i$ and $d_i$ are the $i$-loop coefficients multiplying the leading term.
Both limits correspond to the production of a soft particle.
When the particle is a gluon, and the helicity of the hard particle is inherited from the parent,
the tree splitting function is singular ($Q^+ \to q^+g^\pm$,
$G^+ \to g^+g^\pm$ as $w \to 0$ and $G^+ \to g^+ g^+$ as $w \to 1$).   In each of these cases, the one- and two-loop
splitting functions produce the universal constants $a_1$ and $a_2$, indicating that the soft limit is
independent of the particle type and the helicity of the gluon.
In the soft quark (antiquark) limit, there is no polynomial enhancement.

Furthermore, when the helicity of the hard particle is flipped (as in the $Q^+ \to q^+ g^-$ and $G^+ \to g^+ g^-$ splittings as $w \to 1$) 
there is a tree-level suppression factor of at least $(1-w)$.   
In the two-loop results presented in the previous sections, the
finite contributions to the helicity flip processes $Q^+ \to q^+ g^-$, $G^+ \to g^+ g^-$ and $G^+ \to q^+\bar q^+$ splittings contain apparent quadratic, cubic and quadratic
singularities 
as $w \to 1$.  In each case, the singularity is softened by the behaviour of the numerator, such that only the 
  $G^+ \to g^+ g^-$ splitting has a residual singularity of $1/(1-w)$. 
  This is of course annihilated by the tree-level helicity flip suppression factor of $(1-w)^{3/2}$.

All of these limits show that the soft behaviour of the two-loop splitting functions is well behaved and 
consistent with the expected behaviour.

\section{The two-loop soft splitting function}
\label{sec:soft}

In the limit that a gluon becomes soft,
the colour ordered amplitudes factorise in a similar way to 
Eqs.~(\ref{eq:treecol}),(\ref{eq:onecol}) and (\ref{eq:twocol}).
At tree level we have the usual result,
\begin{equation}
\ket{\mathcal{M}^{(0)}_n(a^{\lambda_a},b^{\lambda_b},c^{\lambda_c},\ldots)} 
\stackrel{{b\rightarrow 0}}{\longrightarrow} 
\mathcal{S}^{(0)}(a,b^{\lambda_{b}},c)
\ket{\mathcal{M}^{(0)}_{n-1}(a^{\lambda_a},c^{\lambda_c},\ldots)}
\end{equation}
The tree level soft splitting amplitudes are given by,
\begin{equation}
\mathcal{S}^{(0)}(a,b^+,c) = \frac{\sqrt{2}\la ac \ra}{\la ab \ra \la bc \ra} ,
\qquad \qquad \mathcal{S}^{(0)}(a,b^-,c) = -\frac{\sqrt{2}[ac]}{[ab][bc]}. 
\end{equation}
The soft amplitudes are independent of the helicities and particle types of the neighboring
legs, $a$ and $c$.

The factorisation of one-loop colour ordered amplitudes in the soft limit proceeds as,
\begin{eqnarray}
\ket{\mathcal{M}^{(1)}_n(a^{\lambda_a},b^{\lambda_b},c^{\lambda_c},\ldots)} &\stackrel{{b\rightarrow 0}}{\longrightarrow}&
\phantom{+}\mathcal{S}^{(0)}(a,b^{\lambda_{b}},c)\ket{\mathcal{M}^{(1)}_{n-1}(a^{\lambda_a},c^{\lambda_c},\ldots)}\nonumber \\ 
&& + \mathcal{S}^{(1)}(a,b^{\lambda_{b}},c)\ket{\mathcal{M}^{(0)}_{n-1}(a^{\lambda_a},c^{\lambda_c},\ldots)}.
\end{eqnarray}
The one-loop soft factor has been found, to all orders in $\epsilon$, 
to be~\cite{Bern:split1QCD,Catani:2000pi} where it is expressed in terms of the tree-level soft factor, 
\begin{equation}
	\mathcal{S}^{(1)}(a,b^\pm,c) = r^{(1)}_{soft}(a,b^\pm,c)\,\,\mathcal{S}^{(0)}(a,b^\pm,c),
\end{equation}
where,
\begin{equation}
	r^{(1)}_{soft}(a,b^\pm,c) = -\frac{c_\Gamma}{2\e^2}N\left( \frac{4\pi\mu^2
	(-\sac)}{(-\sab)(-\sbc)} \right)^\epsilon \Gamma(1+\e)\Gamma(1-\e).
\end{equation}

Two-loop amplitudes have a similar soft limit,
\begin{eqnarray}
\ket{\mathcal{M}^{(2)}_n(a^{\lambda_a},b^{\lambda_b},c^{\lambda_c},\ldots)} &\stackrel{{b\rightarrow 0}}{\longrightarrow}&
\phantom{+}\mathcal{S}^{(0)}(a,b^{\lambda_{b}},c)\ket{\mathcal{M}^{(2)}_{n-1}(a^{\lambda_a},c^{\lambda_c},\ldots)}\nonumber \\ 
&& + \mathcal{S}^{(1)}(a,b^{\lambda_{b}},c)\ket{\mathcal{M}^{(1)}_{n-1}(a^{\lambda_a},c^{\lambda_c},\ldots)}\nonumber \\
&& + \mathcal{S}^{(2)}(a,b^{\lambda_{b}},c)\ket{\mathcal{M}^{(0)}_{n-1}(a^{\lambda_a},c^{\lambda_c},\ldots)}.
\end{eqnarray}
As in the one-loop case, it is convenient to factor out the overall singularity as the tree soft factor,
\begin{equation}
	\mathcal{S}^{(2)}(a,b^\pm,c) = r^{(2)}_{soft}(a,b^\pm,c)\,\,\mathcal{S}^{(0)}(a,b^\pm,c).
\end{equation}

We can use the two-loop quark-gluon and gluon-gluon splitting amplitudes presented in Secs.~\ref{sec:quarkgluon} and \ref{sec:gluongluon}
to recover the soft limit. Compared to the collinear limit defined by Eq.~(\ref{eq:colmom}), 
the soft limit is obtained by taking $w$ small. 
In terms of invariants, $w \to s_{bc}/s_{ab}$.  Expanding the explicit expressions for 
$r^{G\to gg,(2)}(-,g^+,g^+)$ , $r^{G\to gg,(2)}(-,g^+,g^-)$ , $r^{Q\to qg,(2)}(\mp,q^\pm,g^\pm)$ and 
$r^{Q\to qg,(2)}(\mp,q^\pm,g^\mp)$ in the $w \to 0$ limit, we find the universal two-loop soft splitting factor to be,
\begin{eqnarray}
\label{eq:r2soft}
\lefteqn{r^{(2)}_{soft}(a,b^\pm,c) 
=c_\Gamma^2(\e)\frac{\Gamma(1+\e)^2\Gamma(1-\e)^2}{8\e^4}   
\left( \frac{4\pi\mu^2 (-\sac)}{(-\sab)(-\sbc)} \right)^{2\e}} \nonumber \\
&\times& \Bigg( 
N^2-N\beta_0 \e -N K \e^2 +\left[N^2 \left(-\frac{193}{27}+\zeta_3\right) + N\NF \frac{19}{27}\right]\e^3 \nonumber \\
&&
\phantom{+}+\left[N^2 \left(-\frac{1142}{81}+\frac{88}{3}\zeta_3+\frac{7}{180}\pi^4\right) +
N\NF\left(\frac{65}{81}-\frac{16}{3}\zeta_3\right)\right]\e^4
+ {\cal O}(\e^5)\Bigg).
\end{eqnarray}

Note that although the soft limit has simple QED-like factorisation properties for
colour ordered amplitudes, the soft limit of the colour dressed amplitudes is rather more complicated.
Nevertheless, the procedure for colour dressing
is straightforward~\cite{Catani:2000pi,Bern:2lsplit}. 

\section{Conclusions}
\label{sec:conclusions}

In this paper we have studied the universal behaviour of two-loop amplitudes in
the limit where two massless external partons become collinear
using explicit calculations of $\gamma^* \to q \bar qg$ and $H \to 3$~partons.
While we work with  
two-loop amplitudes in the specific case, our
results generalise to processes involving more external  particles. Each of the
amplitudes has a trivial colour structure,  so our results
naturally apply to colour stripped (or ordered) amplitudes.

We find that the pole structure of the 
unrenormalised two-loop splitting amplitudes is given by
Eq.~(\ref{eq:r2}).   After renormalisation, this is in agreement with
the pole structure predicted by Catani~\cite{Catani:polestruc} for generic renormalised
two-loop amplitudes. A similar structure has been found for the gluon 
splitting function in Ref.~\cite{Bern:2lsplit}.

The finite remainders for the two independent quark-gluon splitting functions
are given in Eqs.~(\ref{eq:r2qgpp}) and (\ref{eq:r2qgpm}), while similar
results for the three different helicities needed to describe gluon-gluon
splitting  are given in Eqs.~(\ref{eq:r2ggpp}), (\ref{eq:r2ggpm}) and
(\ref{eq:r2ggmm}).  In this latter case, we find complete agreement with the
results obtained in Ref.~\cite{Bern:2lsplit} using a more general approach  
involving the unitarity method of sewing together tree (and one-loop)
amplitudes to directly construct the two-loop splitting functions.  Finally,
the finite two-loop remainder for the one independent quark-antiquark splitting
function is given in Eq.~(\ref{eq:r2qqpp}).

Our results are valid for the time-like splitting of colour
ordered amplitudes in the 't Hooft-Veltman dimensional regularisation
scheme.    The procedures for extending their range of validity into the
space-like region or dressing with colour are
straightforward~\cite{Bern:2lsplit}.

We have studied the limits where one of the split partons is soft and, although
there are apparent additional polynomial singularities as $w \to 1$ in some of
the splitting functions, find that the logarithms and polylogarithms
multiplying the explicit poles conspire to reduce the degree of the singularity
by two powers.   The net result is that the soft splittings which induce a
helicity flip on the hard parton remain suppressed.  

For the processes where the helicity of the hard parton is conserved, we find
that the soft limit of the radiated soft gluon is universal and is independent
of the helicity or particle type of the hard parton.  This universal factor is
the two-loop soft splitting function and we give an explicit expression for it
in Eq.~(\ref{eq:r2soft}).

The splitting amplitudes presented here are another tool in the armoury of perturbative QCD.
They can serve as a check on calculations of two-loop QCD amplitudes with more external
legs.   Together with the triple collinear limits of one-loop integrals~\cite{Catani:triplecoll}
 and the quadruple
collinear limits of tree amplitudes~\cite{DelDuca:treecoll}, 
they may also be useful in rederiving the  recently
calculated~\cite{Moch:nonsinglet,Vogt:singlet}  three-loop contributions to the
Altarelli-Parisi kernels~\cite{Kosower:evolker}.  Finally, they will provide one of the
ingredients necessary for NNNLO computations of jet cross sections. 

\section*{Acknowledgements}
We thank Babis Anastasiou, Zvi Bern, Lance Dixon,  Thomas Gehrmann and Walter Giele for
helpful discussions.
This work was supported in part by the UK Particle Physics and
Astronomy  Research Council and by the EU Fifth Framework Programme `Improving
Human Potential', Research Training Network `Particle Physics Phenomenology  at
High Energy Colliders', contract HPRN-CT-2000-00149.

\appendix

\section{Infrared factorisation of two-loop amplitudes}

\label{app:catani}

Catani~\cite{Catani:polestruc} has shown how to organize the 
infrared pole structure of the two-loop amplitudes renormalized in the 
\MSbar\ scheme in terms of the tree and renormalized one-loop amplitudes.
In particular, the infrared behaviour of the one-loop coefficients is given by
\begin{equation}
\label{eq:cat1}
|{\cal M}_{\P}^{(1)}\rangle = 
{\bom I}_{\P}^{(1)}(\epsilon) |{\cal M}_{\P}^{(0)}\rangle 
+ |{\cal M}_{\P}^{(1),{\rm finite}}\rangle,\label{eq:IRone}
\end{equation}
where $|{\cal M}_{\P}^{(1)}\rangle$ is the {em renormalised} $i$-loop amplitude. 
Similarly, the two-loop singularity structure is
\begin{equation}
\label{eq:cat2}
|{\cal M}_{\P}^{(2)}\rangle = 
 {\bom I}_{\P}^{(2)}(\epsilon) |{\cal M}_{\P}^{(0)}\rangle  
 + {\bom I}_{\P}^{(1)}(\epsilon) |{\cal M}_{\P}^{(1)}\rangle+ |{\cal M}_{\P}^{(2),{\rm finite}}\rangle,
\label{eq:IRtwo}
\end{equation}
where the infrared singularity operators ${\bom I}_{\P}^{(1)}$ and ${\bom I}_{\P}^{(2)}$ are process but not helicity
dependent. Each of these operators produces terms of ${\cal O}(\epsilon)$ and
the division between finite and non-finite terms is to some extent a matter of
choice. Adding a term of ${\cal O}(1)$ to 
${\bom I}^{(1)}$ is automatically compensated for by a change in ${\bom I}^{(2)}$.

For the virtual photon initiated processes,  (\ref{eq:gamma2}) and 
(\ref{eq:gamma3}),
\begin{eqnarray}
\bom{I}_{\gamma^* \to Q\bar Q}^{(1)}(\epsilon)
&=&
- \frac{e^{\epsilon\gamma}}{2\Gamma(1-\epsilon)} \Biggl[
N \left(\frac{1}{\epsilon^2}+\frac{3}{2\epsilon}\right) 
{\tt S}_{AB}-\frac{1}{N}
\left(\frac{1}{\epsilon^2}+\frac{3}{2\epsilon}\right)
{\tt S}_{AB}\Biggr ], \\
\bom{I}_{\gamma^* \to q\bar q g}^{(1)}(\epsilon)
&=&
- \frac{e^{\epsilon\gamma}}{2\Gamma(1-\epsilon)} \Biggl[
N \left(\frac{1}{\epsilon^2}+\frac{3}{4\epsilon}+\frac{\beta_0}{2N\epsilon}\right) 
\left({\tt S}_{ac}+{\tt S}_{bc}\right)-\frac{1}{N}
\left(\frac{1}{\epsilon^2}+\frac{3}{2\epsilon}\right)
{\tt S}_{ab}\Biggr ],
\end{eqnarray}
where  
\begin{equation}
{\tt S}_{ij} = \left(-\frac{4\pi\mu^2}{s_{ij}}\right)^{\epsilon}.
\end{equation}
Note that on expanding ${\tt S}_{ij}$,
imaginary parts are generated, the sign of which is fixed by the small imaginary
part $+i0$ of $s_{ij}$.

For the Higgs processes,  (\ref{eq:Hgg}), (\ref{eq:Hggg}) and 
(\ref{eq:Hqqg}), we have
\begin{eqnarray}
\bom{I}_{H \to GG}^{(1)}(\epsilon)
&=&
- \frac{e^{\epsilon\gamma}}{2\Gamma(1-\epsilon)} \Biggl[
2N \left(\frac{1}{\epsilon^2}+\frac{\beta_0}{N\epsilon}\right) 
{\tt S}_{AB}\Biggr ], \\
\bom{I}_{H \to ggg}^{(1)}(\epsilon)
&=&
- \frac{e^{\epsilon\gamma}}{2\Gamma(1-\epsilon)} \Biggl[
N \left(\frac{1}{\epsilon^2}+\frac{\beta_0}{N\epsilon}\right) 
\left({\tt S}_{ac}+{\tt S}_{bc}+{\tt S}_{ac}\right)\Biggr ],\\
\bom{I}_{H \to qg\bar q}^{(1)}(\epsilon)
&=&
- \frac{e^{\epsilon\gamma}}{2\Gamma(1-\epsilon)} \Biggl[
N \left(\frac{1}{\epsilon^2}+\frac{3}{4\epsilon}+\frac{\beta_0}{2N\epsilon}\right) \left({\tt S}_{ab}+{\tt S}_{bc}\right) 
-\frac{1}{N}  \left(\frac{1}{\epsilon^2}+\frac{3}{2\epsilon}\right) {\tt S}_{ac} \Biggr ].
\end{eqnarray}

The two-loop singularity operator ${\bom I}_{\P}^{(2)}$ is given by,
\begin{eqnarray}
\lefteqn{{\bom I}_{\P}^{(2)}(\epsilon)}\nonumber \\
&=&
\Biggl (-\frac{1}{2}  {\bom I}_{\P}^{(1)}(\epsilon) {\bom I}_{\P}^{(1)}(\epsilon)
-\frac{\beta_0}{\epsilon} {\bom I}_{\P}^{(1)}(\epsilon) 
+e^{-\epsilon \gamma } \frac{ \Gamma(1-2\epsilon)}{\Gamma(1-\epsilon)} 
\left(\frac{\beta_0}{\epsilon} + K\right)
{\bom I}_{\P}^{(1)}(2\epsilon) + {\bom H}_{\P}^{(2)}(\epsilon) 
\Biggr ).
\end{eqnarray}

In principle, ${\bom H}_{\P}^{(2)}$ contains colour correlations between the final state particles.
However, for the processes at hand, these correlations vanish and 
each external coloured
leg in the  partonic process contributes independently 
to ${\bom H}_{\P}^{(2)}$
\begin{eqnarray}
{\bom H}_{\P}^{(2)}  = \frac{e^{\epsilon\gamma}}{4 \epsilon \Gamma(1-\epsilon)}
\left(n_q H_{q}^{(2)} + n_g H_{g}^{(2)}\right),
\end{eqnarray}
where $n_g$ is the number of external gluons, $n_q$ is the number of external quarks and 
anti-quarks for the process $\P$ and $H_q$ and $H_g$ are defined in Eqs.~(\ref{eq:defHq}) and (\ref{eq:defHg})
respectively.

\providecommand{\href}[2]{#2}\begingroup\raggedright\endgroup
\end{document}